\documentclass[opre]{informs3}
\RequirePackage[colorlinks]{hyperref}

\usepackage{verbatim}
\usepackage{amsmath,amssymb}
\usepackage{url}

\newcommand{\beql}[1]{\begin{equation}\label{#1}}
\newcommand{\eeq}{\end{equation}}

\usepackage[english]{babel}
\usepackage{natbib}
 \bibpunct[, ]{(}{)}{,}{a}{}{,}%
 \def\bibfont{\small}%
 \def\newblock{\ }%

\TheoremsNumberedThrough

\ECRepeatTheorems

\EquationsNumberedThrough

\begin{document}

\setlength{\pdfpageheight}{\paperheight}
\setlength{\pdfpagewidth}{\paperwidth}

\RUNAUTHOR{Cont, Larrard}

\RUNTITLE{Order book dynamics in liquid markets} \TITLE{Order book
dynamics in liquid markets:\\ limit theorems and diffusion approximations}
\ARTICLEAUTHORS{%
\AUTHOR{Rama CONT \& Adrien de LARRARD}
  \AFF{Columbia University, New York\\ \&\\  Laboratoire de Probabilit\'es et Mod\`eles Al\'eatoires\\ CNRS - Universit\'e Pierre et Marie Curie (Paris VI)}
\EMAIL{Revised  Feb 2012}}


\ABSTRACT{ We propose a model for the dynamics of a limit order book in a liquid market where buy and sell orders are submitted at high frequency. We derive a functional central limit theorem for the joint dynamics of the bid and ask queues and show that, when the frequency of order arrivals is large, the intraday dynamics of the limit order book may be approximated by a Markovian jump-diffusion process in the positive orthant, whose characteristics are explicitly described in terms of the statistical properties of the underlying order flow. This result allows to obtain tractable analytical approximations for various quantities of interest, such as the probability of a price increase or the distribution of the duration until the next price move, conditional on the state of the order book. Our results allow for a wide range of distributional assumptions and temporal dependence in the order flow and apply to a wide class of stochastic models proposed for order book dynamics, including models based on Poisson point processes, self-exciting point processes and models of the ACD-GARCH family.
}%
\KEYWORDS{limit order book, queueing systems,  heavy traffic limit, functional central limit
theorem, diffusion limit, high-frequency data, market
microstructure, point process, limit order market}

\maketitle

\newpage

\tableofcontents\newpage

\section{Introduction}\label{sec:intro}

An increasing proportion of financial transactions -in stocks, futures and other contracts- take place in   electronic markets where participants may submit limit orders (for buying or selling), market orders and order cancelations which are then centralized in a {\it limit order book} and executed according to precise time and price priority rules. The limit order book  represents, at each point in time, the outstanding  orders which are awaiting execution: it  consists in queues at different price levels where these orders are arranged according to time of arrival. A limit new buy (resp. sell) order of size $x$ increases
the size of the bid (resp. ask) queue by $x$. Market orders are executed against limit orders at the best available price: a market order decreases of size $x$ the corresponding queue size by $x$. Limit orders placed at the best available price are  executed against market orders.

The availability of high-frequency data on limit order books has generated a lot of interest in statistical modeling of order book dynamics, motivated either by high-frequency trading applications or simply a better understanding of intraday price dynamics (see \cite{cont2011} for a recent survey). The challenge here is to develop statistical models which capture salient features of the data while allowing for some analytical and computational tractability.

Given the discrete nature of order submissions and precise priority rules for their execution, is quite natural to model a limit order book  as a queueing system; early work in this direction dates back to \cite{mendelson82}. More recently,  Cont, Stoikov and Talreja \cite{cont2010} have studied a Markovian queueing model of a limit order book, in which arrivals of market orders and limit orders at each price level are modeled as independent Poisson processes. \cite{contlarrard2010a} used this Markovian queueing approach to compute useful quantities (the distribution of the duration between price changes, the distribution and autocorrelation of price changes, and the probability of an upward move in the price, conditional on the state of the order book) and relate the volatility of the price with statistical properties of the order flow.

\begin{figure}[tbh]
\centering
\includegraphics[width=0.5\textwidth]{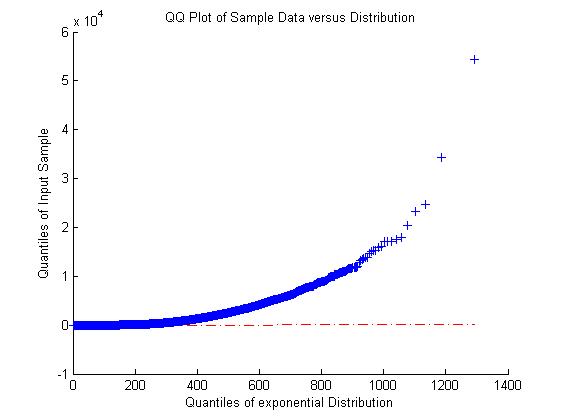}
\caption{Quantiles of inter-event durations compared with quantiles
of an exponential distribution with the same mean (Citigroup, June
2008). The dotted line represents the benchmark case where the
observations are exponentially distributed, which is clearly not the
case here.} \label{fig.qqplotdurations}
\end{figure}

\begin{figure}[tbh]
\centering
\includegraphics[width=9cm]{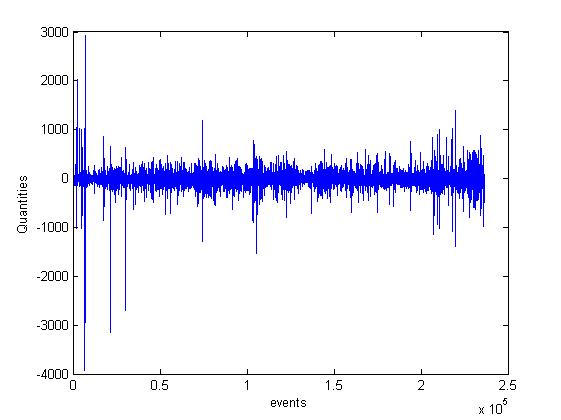}
\caption{Number of shares per event for events affecting the ask. The stock is Citigroup on the 26th of June 2008}
\label{Quantities.chap3}
\end{figure}

However, the results obtained in such Markovian models rely on the fact that
time intervals between orders are independent and exponentially distributed, orders are of the same size
and that the order flow at the  bid is independent from the order flow at the ask.
Empirical studies on high-frequency data  show  these assumptions to be incorrect (\cite{hasbrouck,bouchaud1,bouchaud08,acv2010}). Figure \ref{fig.qqplotdurations} compares the quantiles of the duration between order book events for CitiGroup stock on June 26, 2008 to those of an exponential distribution with the same mean, showing that the empirical distribution of durations is far from being exponential. Figure \ref{fig.acf} shows the autocorrelation function of the inverse durations: the persistent positive value of this autocorrelation shows that durations may not be assumed to be independent.
Finally, as shown in Figure \ref{Quantities.chap3} which displays the (positive or negative) changes in queue size induced by successive orders for CitiGroup shares, there is considerable heterogeneity in  sizes and clustering in the timing of orders.

Other, more complex, statistical models for order book dynamics have been developed to take these properties into account (see Section \ref{sec.examples}). However, only models based on Poisson point processes such as \cite{cont2010,contlarrard2010a}  have offered so far
the analytical tractability necessary  when it comes to studying   quantities  of interest such as durations or transition probabilities of the price, conditional on  the state of the order book.
It is therefore of interest to know whether the conclusions based on  Markovian models are robust to a departure from these simplifying assumptions and, if not, how they must be modified in the presence of other distributional features and dependence
in durations and order sizes.

The goal of this work is to show that it is indeed possible to restore analytical tractability without imposing restrictive assumptions on the order arrival process, by exploiting the {\it separation of time scales} involved in the problem. The existence of widely different time scales, from milliseconds to minutes, makes it possible to obtain meaningful results from an asymptotic analysis of order book dynamics using a {\it diffusion approximation} of the limit order book.  We argue that this diffusion approximation provides relevant and computationally tractable approximations  of the quantities of interest in liquid markets where order arrivals are frequent.

\begin{table}[h]\label{timescales.table}
\begin{center}
\hspace{0.5cm}
\begin{tabular}{|l|c|c|}
\hline Regime  & Time scale & Issues \\ \hline
Ultra-high   & $\sim 10^{-3}-0.1$ s & Microstructure,  \\
frequency (UHF)  & &  Latency  \\ \hline
High   &  $\sim 1-100$ s &  Trade \\
  Frequency (HF)  &     & execution \\ \hline
``Daily"   & $\sim  10^3-10^4$ s  & Trading strategies,\\
& &  Option hedging \\ \hline
\end{tabular}
\vspace{0.5cm}
\caption{A hierarchy of time scales.}
\end{center}
\end{table}

\begin{table}[h]
\begin{center}
\hspace{0.5cm}
\begin{tabular}{|l|c|c|}
\hline
   & Average no. of & Price changes\\
    & orders in 10s &  in 1 day \\ \hline
Citigroup  & 4469 & 12499 \\ \hline

General Electric  & 2356 & 7862 \\ \hline

General Motors  & 1275 &  9016 \\ \hline
\end{tabular}
\end{center}
\vspace{0.5cm}
\caption{Average number of orders in  10 seconds and number of price
changes (June 26th,  2008).}
\label{heavytraffic.table}
\end{table}

As shown in Table  \ref{timescales.table}, most applications involve
the behavior of prices over time scales an order of magnitude larger
than the typical inter-event duration: for example, in optimal trade
execution the benchmark is the Volume weighted average price (VWAP)
computed over a period which may range from 10 minutes to a  day: over such time scales much of the
microstructural details of the market are averaged out. Second, as
noted in Table  \ref{heavytraffic.table}, in liquid equity markets
the number of events  affecting the state of the order book over
such time scales is quite large, of the order of hundreds or
thousands. The typical duration  $\tau_L$ (resp.
$\tau_M$)  between limit orders (resp. market orders and
cancelations) is typically $0.001-0.01\ll 1$ (in seconds). These
observations show that it is relevant to consider {\it heavy-traffic
limits} in which the rate of arrival of orders is large
for studying the dynamics of order books in liquid markets.

In this limit, the  complex  dynamics of the discrete
queueing system is approximated by a simpler system with a
continuous state space, which can be either described by a system of
ordinary differential equations (in the 'fluid limit', where random
fluctuations in queue size vanish) or a system of stochastic
differential equations (in the 'diffusion limit' where random
fluctuations dominate) (\cite{iglehartwhitt70,harrisonnguyen93,Whitt}). Intuitively,
the fluid limit corresponds to the regime of law of large numbers,
where random fluctuations average out and the limit is described by
average queue size, whereas the diffusion limit corresponds to the
regime of the central limit theorem, where fluctuations in queue
size are asymptotically Gaussian. When order sizes or durations fail to have finite moments of first or second order, other scaling limits may intervene, involving L\'evy processes (see \cite{Whitt}) or fractional Brownian motion \cite{araman2011}. As shown by \cite{dai94}, there are also cases where such a 'heavy traffic limit' may fail to exist.
The relevance of each of these
asymptotic regimes is, of course, not a matter of `taste'  but an
empirical question which depends on the behavior of high-frequency order flow in these markets.

Using empirical data on US stocks, we argue that for most liquid
stocks, while the rate of arrival of market orders and limit orders
is large, the {\it imbalance} between limit orders, which increase
queue size, and market orders and cancels, which decrease queue
size, is an order of magnitude smaller: over, say, a 10 minute
interval, one observes an imbalance ranging from 1 to 10 \% of order
flow. In other words, over a time scale of  several minutes, a large
number $N$ of events occur, but the bid/ask imbalance accumulating
over the same interval is of order $\sqrt{N}\ll N$. In this regime,
 random fluctuations in queue sizes cannot be ignored and it is relevant to
consider the {\it diffusion limit} of the limit order book.

In this paper we study the behavior of a limit order book in this
diffusion limit:   we prove a functional central limit theorem for
the joint dynamics of the bid and ask queues when the intensity of
orders becomes large, and use it to derive an analytically tractable
jump-diffusion approximation. More precisely, we show that under a wide range of assumptions, which are shown to be plausible for empirical data on liquid US stocks,  the intraday dynamics of the limit
order book behaves like as a planar Brownian motion in the interior of the positive orthant, and jumps to the interior of the orthant at each hitting time of the boundary.

This jump-diffusion approximation  allows various
quantities of interest to be computed analytically: we  obtain analytical
expressions for various quantities such as
 the probability that the price will increase at the next price
change, and
 the distribution of the duration between price changes,
conditional on the state of the order book.

Our results extend previous analysis of heavy traffic limits for such
auction processes (\cite{kruk03,Bayraktar06,contlarrard2010a}) to a
setting which is relevant and useful for quantitative modeling of
limit order books and provide a foundation for recently proposed
diffusion models for order book dynamics \cite{avellaneda2010}.

\hspace{1cm}

\textbf{Outline.} The paper is organized as follows. Section
\ref{sec.model} describes a general framework for the dynamics of a
limit order book; various examples of models studied in the literature are shown to fall within this modeling framework (Section \ref{sec.examples}).
Section \ref{sec.empirique} reviews some statistical properties of high frequency order flow in limit order markets: these properties
highlight the complex nature of the order flow and motivate the statistical assumptions used to derive the diffusion limit.
Section \ref{sec.heavytraffic}  contains our main result: Theorem \ref{HeavyTraffic.thm} shows that, in a limit order market where orders arrive at high frequency, the bid and ask queues behaves like  a Markov process in the positive quadrant which diffuses inside the quadrant and jumps to the interior each time it hits the boundary.
We provide a complete  description of this process, and use  it to derive,
 in Section
\ref{sec.Markovapproximation}, a simple
jump-diffusion approximation for the joint dynamics of bid and ask
queues, which is  easier to study and simulate
than the initial queueing system.

In particular, we show that in this asymptotic regime the price process is characterized  as a piecewise constant process whose transition times correspond to hitting times of the axes by a two dimensional Brownian motion in the positive orthant (Proposition \ref{prop.pricedynamics}).
This result allows to study analytically various quantities of interest, such as the distribution of the duration between
price moves and the probability of an increase in the price: this is discussed in Section  \ref{sec.computations}.

\section{A   model for the dynamics of a limit order book}
\label{sec.model}
\subsection{Reduced-form representation of a limit order book} \label{secModel}

\begin{figure}[tbh]
\begin{center}
\includegraphics[width=9cm]{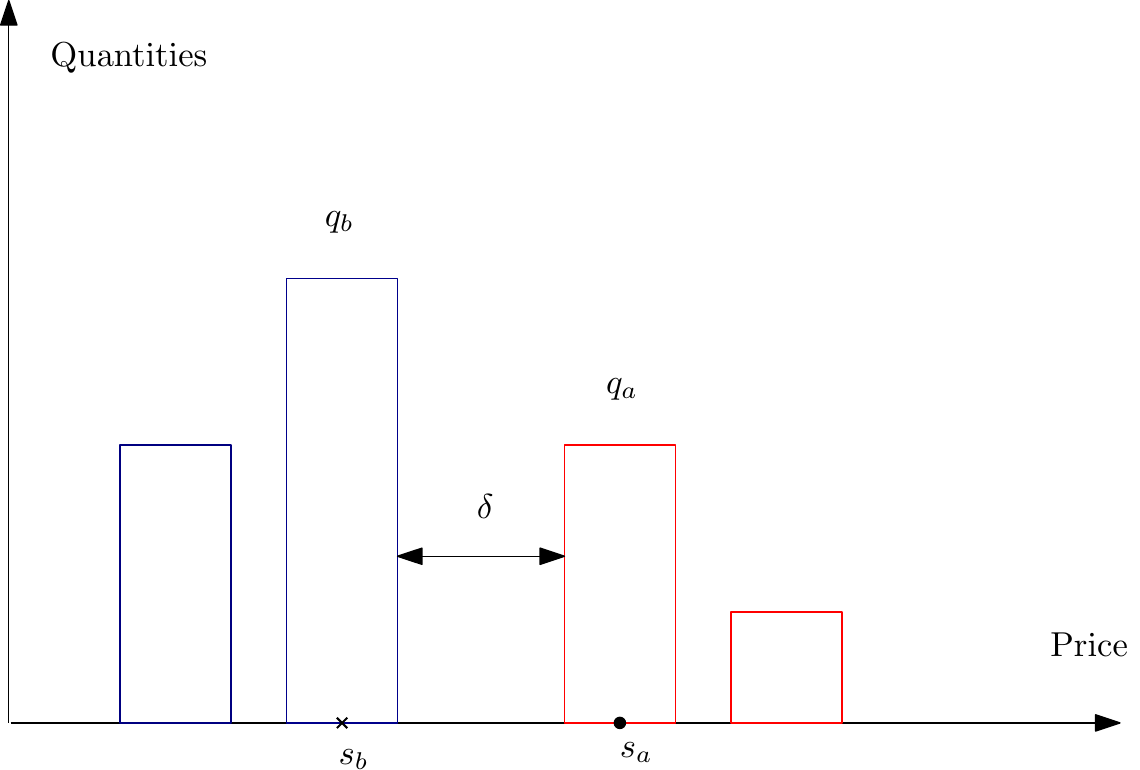}
\caption{Simplified representation of a  limit order book.}
\label{orderbook.fig.ch3}
\end{center}
\end{figure}

Empirical studies of limit order markets
suggest that the major component of the order flow occurs at the (best) bid and ask price levels (see e.g. \cite{biais95}). All electronic trading venues also allow to place limit orders {\it pegged} to the best available price (National Best Bid Offer, or NBBO);  market makers used these \textit{pegged} orders to liquidate their inventories. Furthermore, studies on
the price impact of order book events show that
the net effect of orders on the bid and ask queue sizes is the main factor driving price variations (\cite{cks2010}).
These observations, together with the fact that queue sizes at the best bid and ask of the consolidated order book are
 more easily obtainable (from records on trades and quotes)  than information on deeper levels of the order book, motivate a
reduced-form modeling approach in which we represent the state of the limit order book
by
\begin{itemize}\item the bid price $s_t^b$ and the ask price $s_t^a$
\item the size of the bid queue  $q_t^b$ representing the outstanding limit buy orders at the bid, and
\item the size of the ask queue $q_t^a$  representing the outstanding limit sell orders at the ask
\end{itemize}
Figure \ref{orderbook.fig.ch3}
summarizes this representation.

If the stock is traded in several venues, the quantities $q^{b}$ and $q^{a}$ represent the best bids and offers in the {\it consolidated} order book, obtained by aggregating over all (visible) trading venues. At every time $t$, $q_{t}^{b}$ (resp. $q_{t}^{a}$) corresponds to all visible orders available at the bid price $s_{t}^{b}$ (resp. $s_{t}^{a}$) across all exchanges.

The state  of the order book  is modified by {\it order book events}:
limit orders (at the bid or ask), market orders and cancelations (see \cite{cont2010,cks2010,smith2003}).
A limit buy (resp. sell) order of size $x$ increases the size of the bid (resp. ask) queue by $x$, while
a market buy (resp. sell) order decreases the corresponding queue size by $x$.
Cancelation of $x$ orders in a given queue reduces the queue size by $x$.
Given that we are  interested in the queue sizes at the best bid/ask levels, market orders and
cancelations have the same effect on the queue sizes $(q^b_t,q^a_t)$.

The bid and ask prices are multiples of the tick size $\delta$.
When either the bid or ask queue is depleted by market orders and cancelations,
the price moves up or down  to the next level of the order book.
The  price processes $s^b_t,s^a_t$ are thus  piecewise constant processes whose transitions correspond to hitting times of the  axes $\{ (0,y), y>0 \}\cup \{(x,0), x>0\}$
by the process $q_t = (q^{b}_t, q^{a}_t)$.

If the order book contains no `gaps' (empty levels), these price increments are equal to one tick:
 \begin{itemize}
   \item when  the bid queue is depleted, the (bid) price decreases by one tick.
   \item when  the ask queue is depleted, the (ask) price increases by one tick.
 \end{itemize}
If there are  gaps  in the order book, this results in 'jumps' (i.e. variations of more than one tick) in the price dynamics. We will ignore this feature in what follows but it is not hard to generalize our results to include it.

The quantity $s_t^a-s^b_t$ is the {\it bid-ask spread}, which may be one or several ticks.
As shown in Table \ref{spread.table.ch3}, for liquid stocks the    bid-ask spread
is equal to one tick for more than $98\%$ of observations.

\begin{table}[h]
\begin{center}
\hspace{0.5cm}
\begin{tabular}{|l|c|c|c|}
\hline
Bid-ask spread &  1 tick &  2 tick & $\geq$ 3 tick \\ \hline

Citigroup & 98.82 & 1.18 & 0 \\ \hline

General Electric  & 98.80 & 1.18 & 0.02 \\ \hline

General Motors  & 98.71 & 1.15 & 0.14 \\ \hline
\end{tabular}\vspace{0.5cm}
\caption{Percentage of observations with a given  bid-ask spread (June 26th, 2008).}
\label{spread.table.ch3}
\end{center}
\end{table}
When either the bid or ask queue is depleted, the bid-ask spread widens immediately to more than one tick. 
Once the spread has increased, a flow of limit sell (resp. buy) orders quickly fills the gap and the spread reduces again to one tick.
When a limit order is placed inside the spread, all the limit orders  pegged  to the  NBBO price move in less than a millisecond to the price level corresponding to this new order.
Once this happens, both the bid price and the ask price have increased (resp. decreased) by one tick.

\begin{figure}[tbh]
\begin{center}
\includegraphics[width=0.45\textwidth]{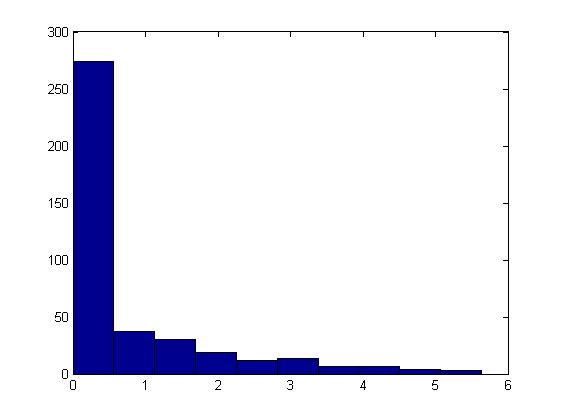}
\includegraphics[width=0.45\textwidth]{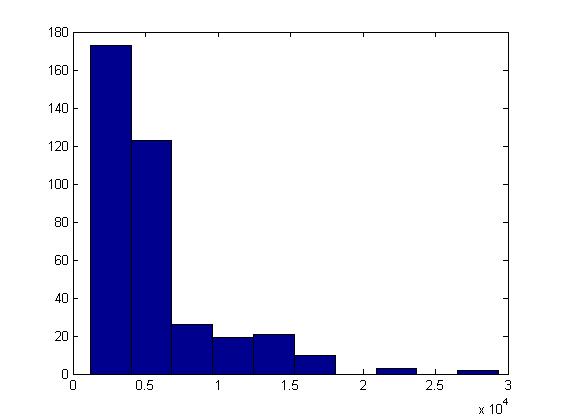}
\caption{Left: Average lifetime, in milliseconds of a spread larger than one tick for Dow Jones stocks. Right: Average lifetime, in milliseconds of a spread equal to one tick.}
\label{tempsfermeture.ch3}
\end{center}
\end{figure}

The histograms in Figure \ref{tempsfermeture.ch3}  show that this 'closing' of the spread takes place very quickly:
as shown in Figure \ref{tempsfermeture.ch3} (left)
  the lifetime of a  spread larger than one tick is of the order of a couple of milliseconds, which  is negligible compared to the lifetime of a spread equal to one tick (Figure \ref{tempsfermeture.ch3} , right).
  In our model we assume that the second step occurs {\it infinitely fast}: once the bid-ask spread widens, it returns immediately to one tick. For the example of Dow Jones stocks (Figure \ref{tempsfermeture.ch3} ),  this is a reasonable assumption since the widening of the spread lasts only a few milliseconds.
This simply means that  we are not trying to describe/model how the orders flow inside the bid-ask spread at the millisecond time scale and, when we describe the state of the order book {\it after } a price change we have in mind the state of the order book {\it after} the bid-ask spread  has returned to one tick.

 Under this assumption, each time {\it one } of the queues is depleted, {\it both} the bid queue  and the ask queues move to a
new position and the bid-ask spread remains equal to one tick after
the price change. Thus, under our assumptions the bid-ask spread is
equal to one tick, i.e. $s_t^a=s_t^b+\delta$, resulting in a further
reduction of dimension in the model.

Once either the bid or the ask queue are depleted, the bid and ask
queues assume new values.  Instead of keeping track of arrival,
cancelation and execution of orders at {\it all} price levels (as in
\cite{cont2010,smith2003}), we treat the queue sizes after a price change
as a stationary sequence of random  variables whose distribution represents the depth of the order book in a statistical sense.
More specifically, we model the size of the bid and ask queues after a price
increase  by a stationary sequence  $(R_k)_{k\geq 1}$ of random  variables with values in $\mathbb{N}^2$.
Similarly, the size of the bid and ask queues after a price
decrease is modeled  by a stationary sequence  $(\tilde{R}_k)_{k\geq 1}$ of random  variables with values in $\mathbb{N}^2$.
 The sequences $(R_k)_{k\geq 1}$ and $(\tilde{R}_k)_{k\geq 1}$ summarize the interaction of the queues at the best bid/ask levels with the rest of the order book,
viewed here as a 'reservoir' of limit orders.

\begin{figure}[tbh]     \centering     \includegraphics[width=0.7\textwidth]{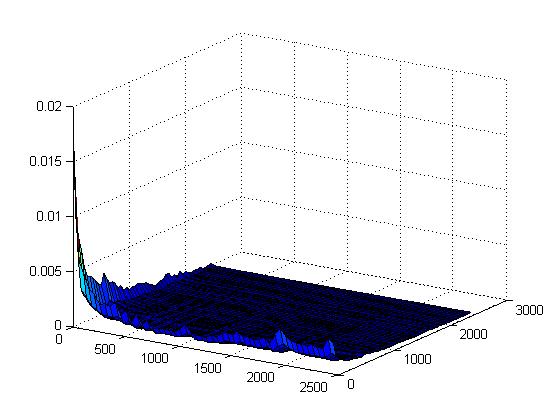}
\caption{Joint density of bid and ask queue sizes after a price move (Citigroup, June 26th 2008).}
\label{f.fig}
\end{figure}

The variables $R_k$ (resp. $\tilde{R}_k$) have a common distribution
 which represents the {\it depth} of the order book after a price
increase (resp. decrease):
Figure \ref{f.fig} shows the (joint) empirical distribution of bid and ask queue sizes after a price move for Citigroup stock on June 26th 2008.

The simplest specification could be to take   $(R_k)_{k\geq 1}$,
$(\tilde{R}_k)_{k\geq 1}$ to be IID sequences;
 this approach, used in \cite{contlarrard2010a},
turns out to be good enough for many purposes.
 But this IID assumption is not necessary; in the next section we will see more general specifications which allow for serial dependence.

In summary,  state of the limit order book is thus described  by a
continuous-time process $(s_t^b,q_t^b,q_t^a)$ which takes values in
the discrete state space $\delta \mathbb{Z} \times  \mathbb{N}^{2}$,
with piecewise constant  sample paths whose transitions correspond
to the order book events. Denoting by $(t_{i}^{a},i\geq 1)$ (resp.
$t_{i}^{b}$) the event times at the ask (resp.   the bid),
$V_{i}^{a}$ (resp. $V_{i}^{b}$) the corresponding change in ask
(resp. bid) queue size, and $k(t)$  the number of price changes in
$[0,t]$,
the above assumptions translate into the following dynamics
for $(s_t^b,q_t^b,q_t^a)$:
\begin{itemize}
\item If an order or cancelation of size $V_{i}^{a}$ arrives on the ask side at $t= t_i^a,$
\begin{itemize}
\item if $q_{t-}^{a}  + V_{i}^{a} >0$, the order can be satisfied without changing the price;
\item if $q_{t-}^{a}  + V_{i}^{a}\leq 0$, the ask queue is depleted, the price increases by one 'tick' of size $\delta$, and the queue sizes take
new values $R_{k(t)}=(R_{k(t)}^{b},R_{k(t)}^{a} )$,
\end{itemize}
\begin{equation}(s_{t}^{b},q_{t}^{b},q_{t}^{a}) = (s_{t-}^{b},q_{t-}^{b},q_{t-}^{a} + V_{i}^{a}) 1_{q_{t-}^{a} > -V_{i}^{a}} + (s_{t-}^{b}+\delta,R_{k(t)}^{b},R_{k(t)}^{a}) 1_{q_{t-}^{a} \leq -V_{i}^{a}}, \label{eq.increase}\end{equation}
\item If an order or cancelation of size $V_{i}^{b}$ arrives on the bid side at $t= t_i^b,$
\begin{itemize}
\item if $q_{t-}^{b}  + V_{i}^{b} >0$, the order can be satisfied without changing the price;
\item if $q_{t-}^{b}  + V_{i}^{b}\leq 0$, the bid queue gets depleted, the price decreases by one 'tick' of
size $\delta$ and the queue sizes take  new values
$\tilde{R}_{k(t)}=(\tilde{R}_{k(t)}^{b},\tilde{R}_{k(t)}^{a})$:
\end{itemize}
\begin{equation}(s_{t}^{b},q_{t}^{b},q_{t}^{a}) = (s_{t-}^{b},q_{t-}^{b} + V_{i}^{b},q_{t-}^{a}) 1_{q_{t-}^{b} > -V_{i}^{b}} + (s_{t-}^{b}-\delta,\tilde{R}_{k(t)}^{b},\tilde{R}_{k(t)}^{a}) 1_{q_{t-}^{b} \leq -V_{i}^{b}}. \label{eq.decrease}\end{equation}
\end{itemize}

\subsection{The limit order book as a 'regulated' process in the orthant}
As in the case of reflected processes arising in queueing networks,
the process $q_t=(q_t^b,q_t^a)$ may  be constructed from the {\bf
net order flow process}  $$x_t=(x^b_t,x^a_t)=\left( \sum_{i=1}^{N^b_t}
V_i^b, \sum_{i=1}^{N^a_t} V_i^a \right)$$ where $N^b_t$ (resp. $N^a_t$) is
the number of events (i.e. orders or cancelations) occurring at the
bid (resp. the ask) during $[0,t]$. $x_t=(x^b_t,x^a_t)$ is analogous
to the 'net input' process in queuing systems \cite{Whitt}:
  $x^b_t$ (resp. $x^a_t$) represents the cumulative
  sum of all orders and cancelations at the bid (resp. the ask) between $0$ and $t$.

$q=(q_t^b,q^a_t)_{t\geq 0}$ which takes values in the positive orthant, may be constructed
from $x$ by  reinitializing its value to a a new position  inside
the positive orthant according to the rules
\eqref{eq.increase}--\eqref{eq.decrease} each time one of the queues
is depleted:  every time $(q_t)_{t\geq 0}$ attempts to exit the
positive orthant, it jumps to a a new position inside the orthant,
taken from the sequence $(R_n,\tilde{R_n})$.

This construction may be done path by path, as follows:
\begin{definition}\label{def.regularization}
Let  $\omega \in D([0,\infty),\mathbb{R}^{2})$ be a right-continuous
function with left limits (i.e. a cadlag function),
 $R=(R_n)_{n\geq 1}$ and $\tilde{R}=(\tilde{R_n})_{n\geq 1}$ two sequences with values in $\mathbb{R}_{+}^{2}$.
There exists a unique cadlag function  $\Psi(\omega,R,\tilde{R}) \in
D([0,\infty),\mathbb{R}_{+}^{2})$ such that
\begin{itemize}
\item For $t<\tau_{1}$, let $ \Psi(\omega,R,\tilde{R})(t)= \omega(t)$ where \begin{equation*}
    \tau_{1} = \inf \lbrace t \geq 0, \ \omega(t).(1,0) \leq 0 \ \text{or} \ \omega(t).(0,1) \leq 0 \rbrace.
    \end{equation*}
    is the first exit time of $\omega$  from the positive orthant.
\item $\Psi(\omega,R,\tilde{R})(\tau_{1})=R_1  $ if $\Psi(\omega,R,\tilde{R})(\tau_{1}-).(0,1) \leq 0,\
\text{and}\ \Psi(\omega,R,\tilde{R})(\tau_{1})=\tilde{R_1}$ if
$\Psi(\omega,R,\tilde{R})(\tau_{1}-).(1,0) \leq 0$.
\item For $k \geq 1$:\\
$\Psi(\omega,R,\tilde{R})(t + \tau_{k}) =
\Psi(\omega,R,\tilde{R})(\tau_{k})+ \omega(t + \tau_{k}) -
\omega(\tau_{k})$ for $0 \leq t < \tau_{k+1} - \tau_{k}$,
where
\begin{equation*} \label{jump.time}
    \tau_{k} = \inf \lbrace t \geq \tau_{k-1} , \ \omega(t).(1,0) \leq \Psi(\omega,R,\tilde{R})(\tau_{k-1}).(1,0) \ \text{or} \ \omega(t).(0,1) \leq \Psi(\omega,R,\tilde{R})(\tau_{k-1}).(0,1) \rbrace
    \end{equation*}
\item  $\Psi(\omega,R,\tilde{R})(\tau_{k})=R_k $  if $\Psi(\omega,R,\tilde{R})(\tau_{k}-).(0,1) \leq \Psi(\omega,R,\tilde{R})(\tau_{k-1}).(1,0)$ and
$\Psi(\omega,R,\tilde{R})(\tau_{k})=\tilde{R_k}$ otherwise.
\end{itemize}\end{definition}

The path $\Psi(\omega,R, \tilde{R})$ is   obtained by "regulating"
the path $\omega$ with the sequences $(R, \tilde{R})$: in between
two exit times, the increments of $\Psi(\omega,R, \tilde{R})$ follow
those of $\omega$ and each  time the process attempts to exit the
positive orthant by crossing the $x$-axis (resp. the $y$-axis), it jumps to a a new position inside the orthant,
taken from the sequence $(R_n)_{n\geq 1}$ (resp. from the sequence $(\tilde{R_n})_{n\geq 1}$).

Unlike the more familiar case of a continuous reflection at the
boundary, which arises in  heavy-traffic limits of multiclass
queueing systems (see \cite{harrison78,harrisonnguyen93,Whitt,ramanan03} for examples), this construction
introduces a discontinuity by pushing the process into the interior
of the positive orthant each time it attempts to exit from the axes.

To study the continuity properties of this map, we endow $D([0,\infty),\mathbb{R}^2)$ with Skorokhod's $J_1$  topology \cite{Billingsley,lindvall73} and the set $(\mathbb{R}_+^2)^\mathbb{^N}$  with
the topology induced by 'cylindrical' semi-norms, defined as follows: for a sequence $(R^n)_{n\geq 1}$ in $(\mathbb{R}_+^2)^\mathbb{^N}$
$$ R^n \mathop{\to}^{n\to\infty} R\in(\mathbb{R}_+^2)^\mathbb{^N}\quad \iff \left(\forall k\geq 1,\quad \sup\{|R^n_1-R_1|,...,|R^n_k-R_k|)\mathop{\to}^{n\to\infty} 0\right).$$
$D([0,\infty),\mathbb{R}^2)\times (\mathbb{R}_+^2)^\mathbb{^N} \times (\mathbb{R}_+^2)^\mathbb{^N}$ is then endowed with the corresponding product topology.
\begin{theorem}\label{prop.psicontinuity}
Let $R=(R_n)_{n\geq 1},\tilde{R}=(\tilde{R}_n)_{n\geq 1}$ be sequences in $]0,\infty[\times ]0,\infty[$ which do not have any
accumulation point on the axes. If
$\omega \in C^0([0,\infty),\mathbb{R}^2)$ is such that
\begin{equation}
(0,0)\notin \Psi(\omega,R,\tilde{R})( [0,\infty)\ ).\label{eq.nonintersection}
\end{equation}
Then the map
\begin{eqnarray}
  \Psi: D([0,\infty),\mathbb{R}^2)\times (\mathbb{R}_+^2)^\mathbb{^N} \times (\mathbb{R}_+^2)^\mathbb{^N}& \to & D([0,\infty),\mathbb{R}_+^2)  \label{eq.Psi}
\end{eqnarray}
is continuous at $(\omega,R,\tilde{R})$.
\end{theorem}
Proof: see Section \ref{sec.psicontinuity} in the Appendix.

This construction may be applied to any cadlag stochastic process:
given a cadlag process $X$ with values in  $\mathbb{R}^{2}$ and
(random) sequences $R=(R_n)_{n\geq 1}$ and
$\tilde{R}=(\tilde{R_n})_{n\geq 1}$  with values in
$\mathbb{R}_{+}^{2}$, the process $\Psi(X,R,\tilde{R})$ is a  cadlag
process with values in $\mathbb{R}_{+}^{2}$.

It is easy to see that the order book process $q_t=(q_t^b,q_t^a)$
may be constructed by this procedure:
\begin{lemma} $q= (q^b,q^a)= \Psi(x,R, \tilde{R})$ where
\begin{itemize}
  \item $x_t=(x^b_t,x^a_t)= \left( \sum_{i=1}^{N^b_t} V_i^b, \sum_{i=1}^{N^a_t} V_i^a \right)$  is the { net order flow} at the bid and the ask,
  \item $R=(R_n)_{n\geq 1}$ is the sequence of queue sizes after a price increase, and
  \item   $\tilde{R}=(\tilde{R_n})_{n\geq 1}$ is the sequence of queue sizes after a price decrease.
\end{itemize}
\end{lemma}
One  can  thus build a statistical model for the limit order book by
specifying the joint law of $x$ and of the regulating sequences $(R,
\tilde{R})$. This approach simplifies the study of the  (asymptotic)
properties of $q_t=(q^b_t,q^a_t)$.
\begin{example}[IID reinitializations]\label{example.IID}
The simplest case  is the case where the queue length after each price change is independent from the history
of the order book, as in \cite{contlarrard2010a}.
      $R=(R_n)_{n\geq 1}$
   and $\tilde{R}=(\tilde{R_n})_{n\geq 1}$ are then IID sequences
   with  values in $]0,\infty[^2$. Figure \ref{f.fig} shows an example of such a distribution for a liquid stock.

The law of the process $Q=\Psi(x,R, \tilde{R})$ is then entirely
determined by
 the law of the net order flow $x$ and
  the distributions of $R_n$, $\tilde{R}_n$: it can be constructed from the concatenation of the laws of $(x_{t}, \tau_{k} \leq  t < \tau_{k+1})$ for   $k \geq 0$ (where we define $\tau_{0} := 0$).
\end{example}

 \begin{example}[Pegged limit orders]
 Most electronic trading platforms allow to place limit orders which are {\it pegged} to the best quote: if the best quote moves to a new price level, a pegged limit order moves along with it to the new price level.
 The presence of pegged orders leads to positive autocorrelation and dependence in the queue size before/after a price change.
The queue size after a  price change may be modeled as
\begin{itemize}
  \item $ q_{\tau_n}=R_n=(\epsilon^b_n+ \beta
  q^b_{\tau_n-},\epsilon^a_n)$ if the price has increased, and
  \item $ q_{\tau_n}=\tilde{R_n}=(\tilde{\epsilon}^b_n,\tilde{\epsilon}^a_n+\tilde{\beta}
  q^a_{\tau_n-})$ if the price has decreased
\end{itemize}
where $\epsilon_n=(\epsilon^b_n,\epsilon^a_n),
\tilde{\epsilon}_n=(\tilde{\epsilon}^b_n,\tilde{\epsilon}^a_n) $ are
IID sequences.
 Empirically, one observes a correlation of $\sim
10\%-20\%$ between the queue lengths before and after a price
change, which suggests an order magnitude for the fraction of pegged
orders.

As in the previous example, the law of  of the process $q=\Psi(x,R,
\tilde{R})$ is   determined by
 the law of the net order flow $x$, the coefficients $\beta,\tilde{\beta}$ and
  the distributions of ${\epsilon}$,$\tilde{\epsilon}$: it can be constructed from the concatenation of the laws of $(x_{t}, \tau_{k} \leq  t < \tau_{k+1})$ for   $k \geq 0$.\label{example.pegged}
\end{example}
More generally, one could consider other extensions where the queue size after a price move may depend in a (nonlinear) way on the queue size before the price move and a random term $\epsilon_n$ representing the inflow of new orders after the $n$-th price change:
\begin{equation} q_{\tau_n}= g(q_{\tau_n-}, \epsilon_n).\label{eq.generalRn}\end{equation}
The results given below hold for this general specification although
the examples  \ref{example.IID} and \ref{example.pegged} above are
sufficiently general for most applications.

\subsection{Examples} \label{sec.examples}

The framework described in Section \ref{secModel} is quite general: it allows a wide class of specifications for the order flow
process,and contains as special cases various models proposed in the literature.
Each model involves a specification for the (random) sequences $(t_i^a,t_i^b,V_i^a,V_i^b)_{i\geq 1}$,  $R=(R_n)_{n\geq 1}$ and $\tilde{R}=(\tilde{R_n})_{n\geq 1}$ or, equivalently, $(T_i^a,T_i^b,V_i^a,V_i^b)_{i\geq 1}$,  $R=(R_n)_{n\geq 1}$ and $\tilde{R}=(\tilde{R_n})_{n\geq 1}$ where $T_i^a=t_{i+1}^a-t_i^a$ (resp. $T_i^b=t_{i+1}^b-t_i^b$) are the durations between order book events on the ask (resp. the bid) side.

\subsubsection{Models based on Poisson point processes}
 \cite{contlarrard2010a} study a stylized model of a limit order market in which
 market orders, limit orders and cancelations arrive at independent and exponential times with corresponding rates $\mu$, $\lambda$ and $\theta$, the process $q = (q^{b}, q^{a})$ becomes a Markov process. If we assume additionally that all orders have the same size, the dynamics of the reduced limit order book is described by:
\begin{itemize}

\renewcommand{\labelitemi}{$\bullet$}

\item The sequence $(T_{i}^{a})_{i \geq 0}$ is a sequence of independent random variables with exponential distribution with parameter $\lambda + \theta + \mu,$

\item The sequence $(T_{i}^{b})_{i \geq 0}$ is a sequence of independent random variables with exponential distribution with parameter $\lambda + \theta + \mu,$

\item The sequence $(V_{i}^{a})_{i \geq 0}$ is a sequence of independent random variables with
 $$\mathbb{P}[V_{i}^{a} = 1] = \dfrac{\lambda}{\lambda + \mu + \theta} \ \  and \ \  \mathbb{P}[V_{i}^{a} = -1] = \dfrac{\mu + \theta}{\lambda + \mu + \theta},$$

\item The sequence $(V_{i}^{b})_{i \geq 0}$ is a sequence of independent random variables with
 $$\mathbb{P}[V_{i}^{b} = 1] = \dfrac{\lambda}{\lambda + \mu + \theta} \ \  and \ \  \mathbb{P}[V_{i}^{b} = -1] = \dfrac{\mu + \theta}{\lambda + \mu + \theta}.$$
\item All these sequences are independent.
\end{itemize}
It is readily verified that this model is a special case of the
framework of Section \ref{secModel}: $(q_t)_{t \geq 0}$ may be
constructed as in Definition \ref{def.regularization}, where the
unconstrained process $x_t$ is now a compound Poisson process.

\subsubsection{Self-exciting point processes}

Empirical studies of order durations highlight the dependence in the sequence of order durations.
This feature, which is not captured in models based on  Poisson
processes, may be adequately represented by a multidimensional {\it
self-exciting} point process \cite{acv2010,Hautsch04}, in which the
arrival rate $\lambda_i(t)$ of an order of type $i$
 is represented as a stochastic process
whose value depends on the recent history of the order flow: each
new order increases the rate of arrival for subsequent orders of the
same type (self-exciting property) and may also affect the rate of
arrival of other order types (mutually exciting property):
 $$\lambda_i(t)= \theta_i + \sum_{j=1}^J \delta_{ij} \int_0^t e^{-\kappa_i(t-s)} dN_j(s)$$
Here $\delta_{ij}$ measures the impact of events of type $j$ on the
rate of arrival of subsequent events of type $i$: as each event of
type $j$ occurs, $\lambda_i$ increases by $\delta_{ij}$. In between
events, $\lambda_i(t)$ decays exponentially at rate $\kappa_i$.
Maximum likelihood estimation of this model on TAQ data
\cite{acv2010} shows evidence of self-exciting and mutually exciting
features in order flow: the coefficients $\delta_{ij}$ are all
significantly different from zero and positive, with $\delta_{ii} >
\delta_{ij}$ for $j\neq i$.

\subsubsection{Autoregressive conditional durations}
Models based on Poisson process fail to capture serial dependence in the sequence of durations,
which manifests itself in the form of clustering of order book events.
One approach for incorporating serial dependence in event durations is to represent
 the duration  $T_i$ between transactions $i-1$ and $i$  as
$$ T_{i} = \psi_{i} \epsilon_{i} ,$$
where $(\epsilon_{i})_{i \geq 1}$ is a   sequence of independent positive random variables with common distribution   and
$\mathbb{E}[\epsilon_{i}] = 1$ and the {\it conditional duration  }
$\psi_{i}  = \mathbb{E}[T_{i} | \psi_{i-j}, T_{i-j}, j\geq 1 ]$ is modeled as a function of past history of the process:
$$ \psi_{i} = G(\psi_{i-1}, \psi_{i-2},...,..; T_{i-1}, T_{i-2},...,..). $$
Engle and Russell's Autoregressive Conditional Duration model \cite{engle98}  propose an ARMA$(p,q)$ representation for $G$:
$$ \psi_{i} = a_{0} + \sum_{i=1}^{p} a_{k}\psi_{i-k} + \sum_{i=1}^{q} b_{q}T_{j-k} $$
where $(a_{0},...,a_{p})$ and $(b_{1},...,b_{q})$ are positive
constants. The ACD-GARCH  model  \cite{ghysels98} combine this
model
  with a GARCH model for the returns. \cite{engle00} proposes a GARCH-type model with random durations where the volatility of  a price change may depend on the previous durations. Variants and extensions are discussed in \cite{Hautsch04}.
Such models, like ARMA or GARCH models defined on fixed time intervals, have
likelihood functions which are numerically computable.
Although these references focus on transaction data, the framework can be adapted to
 model the  durations $(T_{i}^{a}, i \geq 1)$ and $(T_{i}^{b}, i \geq 1)$ between  order book events  with the ACD framework  \cite{Hautsch04}.
\subsubsection{A  limit order market with patient and impatient agents}\label{sec.ordersplit}
Another way of specifying a stochastic model for the order flow in a limit order market is to use an 'agent-based' formulation where agent types are characterized in terms of the statistical properties of the order flow they generate. 
Consider for example a market with three  types of traders:
\begin{itemize}
\item {\it impatient} traders who only submit  market orders:
\item  {\it patient} traders who   use \textit{only limit orders}: this is the case for example of traders who place stop loss orders or engage in strategies such as mean-reversion arbitrage or  pairs trading which  are only profitable with limit orders.
\item other  traders who use \textit{both limit and market orders}; we will assume these traders  submit  a proportion $\gamma$ of their orders as limit orders and $(1- \gamma)$ as market orders, where $0 < \gamma < 1$.
\end{itemize}
Denote by $m$ (resp. $l$) the proportion of orders generated by impatient (resp. patient) traders:
\begin{eqnarray}
\forall i \geq 1, \ \  \mathbb{P}[i-\text{th trader uses only market orders}] = m,\nonumber\\
\mathbb{P}[i-\text{th trader uses only limit orders}] =l\nonumber,\\
\mathbb{P}[i\text{th trader uses both limit and market orders}] =1 - l - m.\nonumber
\end{eqnarray}
Assume that the sequence $(T_{i}, i \geq 1)$ of duration between consecutive
orders is a  stationary  ergodic sequence of
random variables with $\mathbb{E}[T_{i}] < \infty$, that each trader has an equal chance of being a buyer or a seller
and that the type of trader (buyer or seller) is independent from
the past:
\begin{equation*}
\mathbb{P}[i-\text{th trader is a buyer}] =  \mathbb{P}[i-\text{th trader is a seller}] = \frac{1}{2}
\end{equation*}
Trader $i$ generates an order of size $V_i$, where $(V_{i}, i \geq 1)$ is an IID sequence with:

\begin{equation*}
\mathbb{P}[(V_{i}^{b},V_{i}^{a}) = (V_{i},0)] = \mathbb{P}[(V_{i}^{b},V_{i}^{a}) = (0,V_{i})] = \frac{m}{2},
\end{equation*}
\begin{equation*}
\mathbb{P}[(V_{i}^{b},V_{i}^{a}) =( -V_{i},0)] = \mathbb{P}[(V_{i}^{b},V_{i}^{a}) =( 0,-V_{i})] = \frac{l}{2},
\end{equation*}
\begin{equation*}
\mathbb{P}[(V_{i}^{b},V_{i}^{a}) = (\gamma V_{i}, -(1- \gamma)V_{i})]= \mathbb{P}[(V_{i}^{b},V_{i}^{a}) = (-(1-\gamma) V_{i}, \gamma V_{i})] = \frac{1 - l - m}{2}.
\end{equation*}
\section{Statistical properties of high-frequency order flow} \label{sec.empirique}
As described in Section \ref{secModel}, the sequence of order book
events --the {\it order flow}-- is characterized by the sequences
$(T_{i}^{a}, i \geq 1)$ and $(T_i^b, i \geq 1)$ of durations between
orders  and the sequences of order sizes $(V_{i}^{b}, i \geq 1)$ and
$(V_{i}^{a}, i \geq 1)$. In this section we illustrate the
statistical properties of these sequences  using
high-frequency quotes and trades for  liquid US stocks
--CitiGroup, General Electric, General Motors-- on June 26th, 2008.

\subsection{Order sizes}

Empirical studies \cite{bouchaud1,bouchaud08,Gopikrishnan,maslov} have shown that order
sizes are   highly heterogeneous and exhibit heavy-tailed
distributions, with Pareto-type tails:
$$ \mathbb{P}(V_i^a \geq x)\sim C x^{-\beta} $$
with tail exponent $\beta>0$ between 2 and 3, which corresponds to a
series with finite variance but infinite moments of order $\geq 3$.
The tail exponent $\beta>0$ is difficult to estimate precisely, but
the Hill estimator \cite{resnick} can be used to measure the
heaviness of the tails. Table \ref{table.durationtail.ch3} gives the Hill estimator of the tail
coefficient of order sizes for our samples. This estimator is larger
than $2$ for both the bid and the ask; this means that the sequence
of order sizes have a finite moment of order two.

\begin{center}
\begin{figure}[h!]
\begin{tabular}{cc}
   \includegraphics[width=8cm]{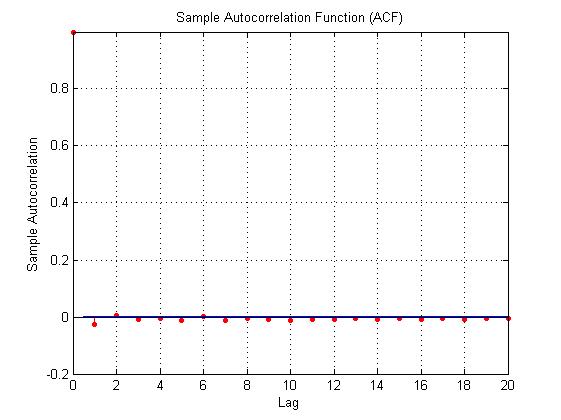} &
   \includegraphics[width=8cm]{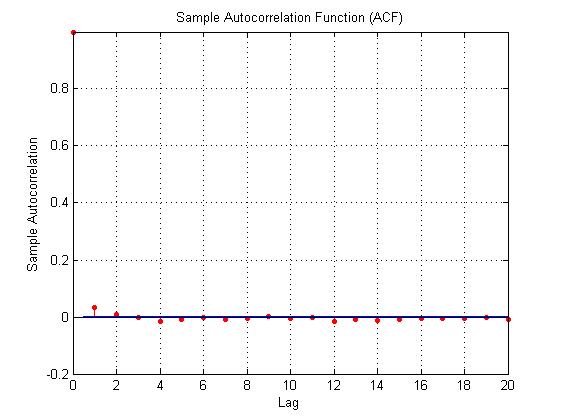} \\
\end{tabular}
\caption{Autocorrelogram of the sequence of order sizes. Order
coming at the ask on the left and at the bid on the right.}
\label{fig.autocorrvol}
\end{figure}

\end{center}

\begin{center}
\begin{figure}[h!]
\begin{tabular}{cc}
   \includegraphics[width=8cm]{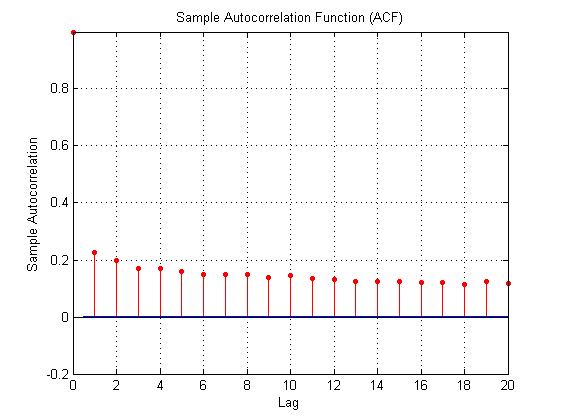} &
   \includegraphics[width=8cm]{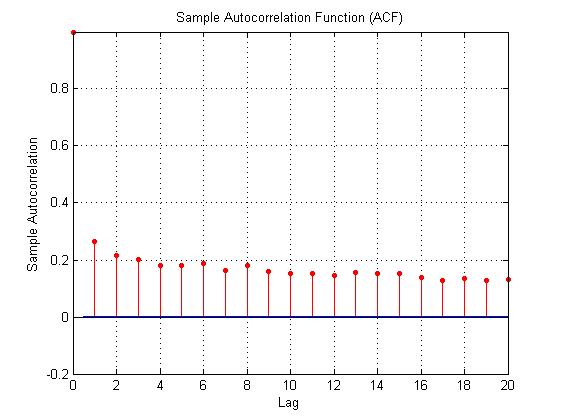} \\
\end{tabular}
\caption{Autocorrelogram of the sequence of absolute order sizes.
Order coming at the ask on the left and at the bid on the
right.}\label{autocorrvolabs.ch3}
\end{figure}
\end{center}

\begin{table}[h]
\begin{center}
\hspace{0.5cm}
\begin{tabular}{|c|c|c|}
\hline
   & Bid side & Ask side \\
   \hline
Citigroup  & [0.42, 0.46] & [0.29, 0.32] \\ \hline

General Electric  & [0.42, 0.45] & [0.41, 0.46] \\ \hline

General Motors  & [0.36, 0.42] & [0.44, 0.51]  \\ \hline
\end{tabular}
\vskip 0.5cm
\caption{$95$-percent confidence interval of the Hill estimator of the sequence of order sizes. When the Hill estimator is  $<0.5$, the estimated  tail index
is large than $ 2$ and the distribution has finite variance. }
\label{table.durationtail.ch3}
\end{center}
\end{table}
 The sequences of order sizes $(V_{i}^{a}, i \geq 1)$ and
$(V_{i}^{b}, i \geq 1)$ exhibit insignificant autocorrelation, as observed
on Figure \ref{fig.autocorrvol}.  However, they are far from being
independent: the series of squared order sizes $((V_{i}^{b})^{2}, i
\geq 1)$ and $((V_{i}^{a})^{2}, i \geq 1)$  are positively
correlated, as revealed by their autocorrelation functions
(displayed in Figure \ref{autocorrvolabs.ch3}).

Finally, the  sequences $(V_{i}^{a}, i \geq 1)$ and $(V_{i}^{b}, i
\geq 1)$ may be negatively correlated. This stems from the fact that a buyer  can simultaneously
use market orders on the ask side (which correspond to negative
values of $V_i^a$ and limit orders on the bid side (which correspond
to positive values of $V_i^b$); the same argument holds for sellers
(see Section \ref{sec.ordersplit}).

These properties of the sequence  $(V_{i}^{a}, V_{i}^{b})_{i \geq
1}$ may be modeled using a bivariate ARCH process:
\begin{align*} \label{Volume.model}
V_{i}^{b}  = \sigma_{i}^{b} z_{i}^{b} \qquad & V_{i}^{a} =
\sigma_{i}^{a} z_{i}^{a}\\
(\sigma_{i}^{b})^{2}  = \alpha^{b}_{0} + \alpha_{1}^{b} (V_{t-1}^{b})^{2},\qquad &
(\sigma_{i}^{a})^{2} = \alpha^{a}_{0} + \alpha_{1}^{a}
(V_{t-1}^{a})^{2},\qquad
{\rm where}\qquad&
(z_{i}^{b},z_{i}^{a})_{i \geq 1}  \mathop{ \sim}^{IID}N \left(0, \begin{pmatrix}
1 & \rho \\
\rho & 1
\end{pmatrix} \right)
\end{align*}
and $(\alpha_{0}^{b},\alpha_{1}^{b},\alpha_{0}^{a},\alpha_{1}^{a})$
are positive coefficients satisfying
\begin{equation}\label{existence}
0< \alpha_{0}^{b} + \alpha_{1}^{b} < 1, \ \ \text{and} \ \ 0< \alpha_{0}^{a} + \alpha_{1}^{a} < 1.
\end{equation}
As shown by \cite{bougerol92}, under the assumption \eqref{existence}, the
sequences of order sizes $(V_{i}^{b}, i \geq 1)$ and $(V_{i}^{a}, i
\geq 1)$ is then a well defined, stationary sequence of random
variables with finite second-order moments, satisfying the
properties enumerated above.
\subsection{Durations} \label{sec.durations}
\begin{center}
\begin{figure}[h!]
\begin{tabular}{cc}
   \includegraphics[width=8cm]{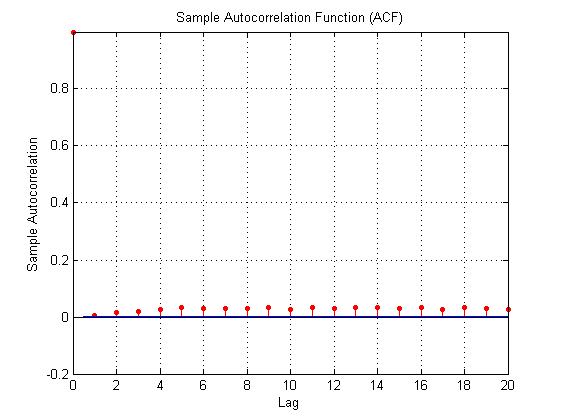} &
   \includegraphics[width=8cm]{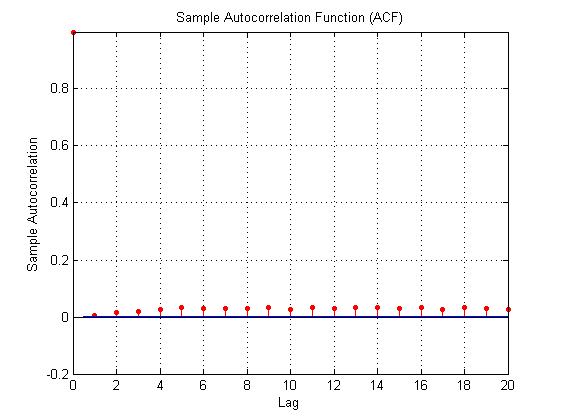} \\
\end{tabular}
\caption{Autocorrelogram of the sequence of   durations for events
at the ask (left) and the bid (right). }
\label{autocorrdur}\end{figure}
\end{center}

\begin{center}
\begin{figure}[h!]
\begin{tabular}{cc}
   \includegraphics[width=8cm]{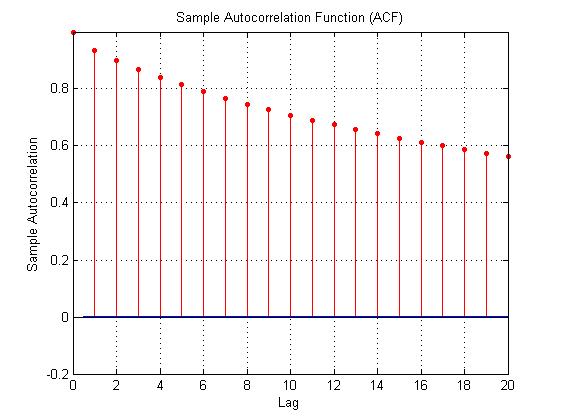} &
   \includegraphics[width=8cm]{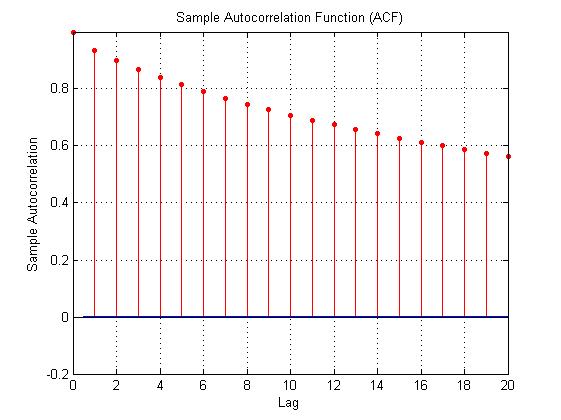} \\
\end{tabular}
\caption{Autocorrelogram of the sequence of inverse durations for
events at the ask (left) and the bid (right).}
\label{fig.acf}
\end{figure}
\end{center}

The timing of order book events is describe by the sequence of durations  $(T_{i}^{b}, i \geq 1)$ at the bid  and $(T_{i}^{a}, i \geq 1)$  at the ask. These sequences have zero autocorrelation
 (see Figure \ref{autocorrdur}) but are not independence sequences: for example, as shown in Figure \ref{fig.acf}, the
sequence of inverse durations $(1/T_{i}^{b}, i \geq 1)$ and
$(1/T_{i}^{b}, i \geq 1)$ is strongly correlated in each case.

\begin{center}
\begin{figure}[h!]
\begin{tabular}{cc}
   \includegraphics[width=8cm]{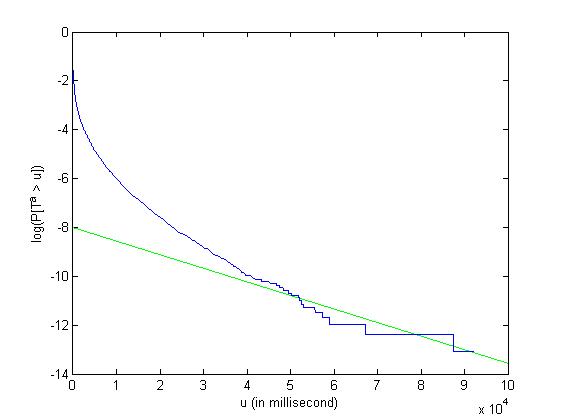} &
   \includegraphics[width=8cm]{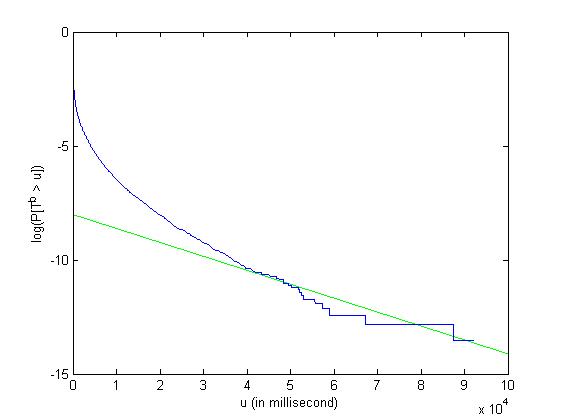} \\
\end{tabular}
\caption{Logarithm of the empirical distribution function of
durations for events at the ask (left) and the bid (right).}
\label{durationtail}
\end{figure}
\end{center}

Figure \ref{durationtail} represents the empirical distribution
functions $ \mathbb{P}[T^{a} > u])$ and $ \mathbb{P}[T^{b} > u]$ in
logarithmic scale. Both empirical distributions exhibit thin,
exponential-type tails (which implies in particular that $T^{a}$ and
$T^{b}$ have finite expectation).

\clearpage

\section{Heavy Traffic limit}\label{sec.heavytraffic}

 At very high frequency, the limit order book is described by a two-dimensional piecewise constant  process $q_t = (q_{t}^{b},q_{t}^{a})_{t \geq 0}$, whose evolution is  determined by the flow of orders. The complex nature of this order flow --heterogeneity and serial dependence in  order sizes,  dependence between orders coming at the ask and at the bid-- described in section \ref{sec.empirique}, makes it difficult to describe $q_t$ in an analytically tractable manner which would allow  the quantities of interest to  be computed either in closed form or numerically in real time applications.
 However, if one is interested in the evolution of the order book over  time scales much larger than the interval between individual order book events,  the (coarse-grained) dynamics of the queue sizes may be described in terms of a simpler process $Q$, called the \textit{heavy traffic approximation} of $q$.
  In this limit, the  complex  dynamics of the discrete
queueing system is approximated by a simpler system with a
continuous state space, which can be either described by a system of
ordinary differential equations (in the 'fluid limit', where random
fluctuations in queue sizes vanish) or a system of stochastic
differential equations (in the 'diffusion limit' where random
fluctuations dominate).
This idea has been widely used in queueing theory to obtain useful analytical insights into the dynamics  of queueing systems    \cite{harrisonnguyen93,iglehartwhitt70,Whitt}.

We argue in this section that the heavy traffic limit is highly  relevant for the study of limit order books in liquid markets, and that the correct scaling limit for the liquid stocks examined in our data sets is the "diffusion" limit. This {\it heavy traffic} limit is then derived in Section \ref{sec.heavytraffic} and described in Section {sec.Markovianapproximation}.

\subsection{Fluid limit or diffusion limit?}\label{sec.fluidornotfluid}

One way of viewing the heavy traffic limit is to view the limit order book at a lower time resolution, by grouping together events in batches of size $n$. Since the inter-event durations are finite, this is equivalent to rescaling time by $n$. The impact, on the net order flow, of a batch of $n$ events at the ask  is
\begin{equation*}
\dfrac{V_{1}^{a} + V_{2}^{a} + V_{3}^{a} + ... + V_{n}^{a}}{\sqrt{n}}  = \dfrac{(V_{1}^{a} - \overline{V^a}) + (V_{2}^{a} - \overline{V^a})+ ... + (V_{n}^{a}- \overline{V^a})}{\sqrt{n}} + \sqrt{n}\quad \overline{V^a},
\end{equation*}
where $(V_{i}^{a}, i \geq 1)$ is the sequence of order sizes   at the ask and $\overline{V^a}=\mathbb{E}[V_{1}^{a}]$. Under appropriate assumptions (see next section), this sum behaves approximately as a Gaussian random variable for large $n$:
\begin{equation}
\dfrac{V_{1}^{a} + V_{2}^{a} + V_{3}^{a} + ... + V_{n}^{a}}{\sqrt{n}}  \sim N(\sqrt{n}\  \overline{V^a}, {\text{Var}(V_{1}^{a})}) \ \  \text{as} \ \ n \rightarrow \infty.
\end{equation}
Two regimes are possible, depending on the behavior of the ratio $\frac{\sqrt{n}\ \overline{V^a}}{\sqrt{\text{Var}(V_{1}^{a})}}$ as $n$ grows:
\begin{itemize}
\item If $\frac{\sqrt{n}\ \overline{V^a}}{\sqrt{\text{Var}(V_{i}^{a})}} \rightarrow \infty$ as $n \rightarrow \infty$, the correct approximation is given by the {\it fluid limit}, which describe the (deterministic) behavior of the average queue size.
\item If $\lim_{n\to \infty } \frac{\sqrt{n} \ \overline{V^a}}{\sqrt{\text{Var}(V_{i}^{a})}}  < \infty$, the rescaled queue sizes behave  like a diffusion process.
\end{itemize}
The fluid limit corresponds to the regime of law of large numbers,
where random fluctuations average out and the limit is described by
average queue size, whereas the diffusion limit corresponds to the
regime of the (functional) central limit theorem, where fluctuations in queue
size are asymptotically Gaussian.

\begin{center}
\begin{figure}[h!]
\begin{tabular}{cc}
   \includegraphics[width=8cm]{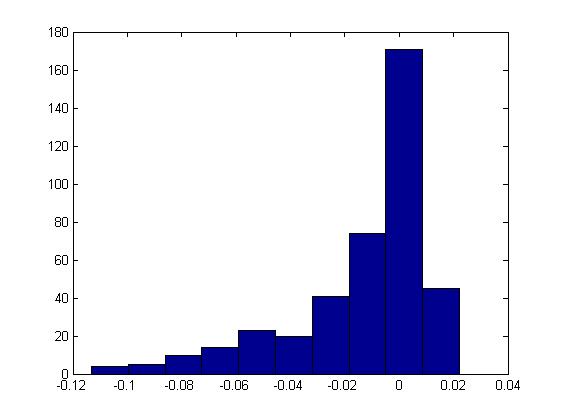} &
   \includegraphics[width=8cm]{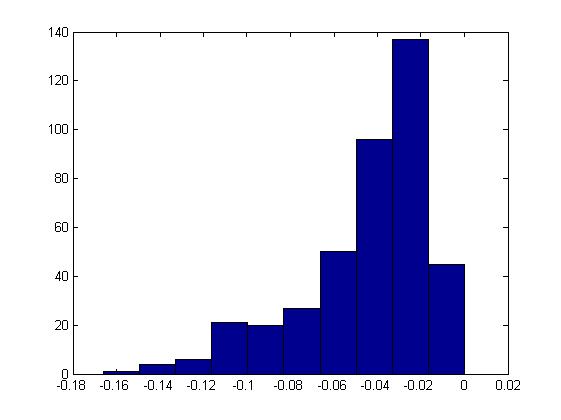} \\
\end{tabular}
\caption{Empirical distribution of the ratio ${\sqrt{n}/ \overline{V^a}}{\sqrt{\text{Var}(V_{1}^{a})}}$ showing the relative importance of average change vs fluctuations in queue size, for stocks in the Dow-Jones index during June 08 (see Section \ref{sec.fluidornotfluid}). Low values of the ratio indicate that intraday changes in bid/ask queue size are dominated by fluctuations, rather than the average motion of the queue.
Left: bid side. Right: ask side.}
\label{voldrift.graph}
\end{figure}
\end{center}

\begin{center}
\begin{figure}[tbh]
\centering
\includegraphics[width=9cm]{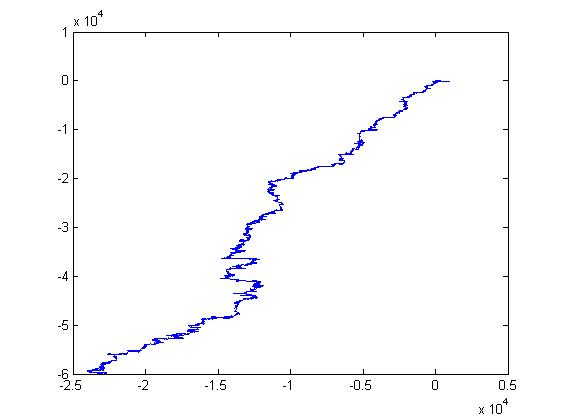}
\caption{Evolution of the net order flow $X_t=(X^b_t,X^a_t)$ given in Eq. \eqref{equation.x} for CitiGroup shares over one trading day (June 26, 2008).}
\label{fig.citibook}
\end{figure}
\end{center}

Figure \ref{voldrift.graph} displays the histogram of the ratio $\frac{\sqrt{n}\ \overline{V^a}}{\sqrt{\text{Var}(V_{i}^{a})}}$
for stocks in the Dow Jones index, where for each stock $n$ is chosen to represent the average number of order book events in a 10 second interval (typically $n\sim 100 -1000$).
This ratio is shown to be rather small at such intraday time scales, showing that the diffusion approximation, rather than the fluid limit, is the relevant approximation to use here.
Indeed,  bid and ask queue sizes $(q_{t}^{b},q_{t}^{a})$ exhibit a diffusion-type behavior at such intraday time scales:
 Figure \ref{fig.citibook} shows   the path of the net   order flow  process
\begin{equation}\label{equation.x}
 x_{t}  =  (q_{0}^{b},q_{0}^{a}) +  \left( \sum_{i=1}^{N_{t}^{b}} V_{i}^{b}, \sum_{i=1}^{N_{t}^{a}} V_{i}^{a}  \right)
\end{equation}
sampled every second for CitiGroup stocks  on a typical trading day.
In this example,  for which the average time between consecutive orders is $\lambda^{-1}\simeq 13\ ms \ll 1$ second,
we observe that   the process $X$ behaves like a diffusion in the orthant with negative drift: the randomness of queue sizes does not average out at this time scale.
We will now show that this is a general result: under mild assumptions on the order flow process, we will show that
the (rescaled) queue size process
\begin{equation}\label{qn}
 (\dfrac{q^{b}_{nt}}{\sqrt{n}},\dfrac{q^{a}_{nt}}{\sqrt{n}})_{ t \geq 0}
\end{equation}
converges in distribution to a Markov process $(Q_t)_{t\geq 0}$ in the positive orthant, whose features we will describe in terms of the statistical properties of the order flow.
\subsection{A functional central limit theorem for the limit order book}\label{sec.fclt}
Consider now a sequence  $q^{n}=(q^{n}_t)_{t\geq 0}$ of processes, where   $q^{n}$ represents the dynamics of the bid and ask queues in the limit order book at a time resolution corresponding to $n$ events (see discussion above).
The dynamics of  $q^{n}$ is characterized by the sequence  of  order sizes $(V_{i}^{n,b},V_{i}^{n,a})_{i \geq 1}$, durations $(T_{i}^{n,b},T_{i}^{n,a})_{i \geq 1}$ between orders and the fact that, at each price change
\begin{itemize}
\item $q^n_{\tau_k}= R^n_k = g(q^n_{\tau_k-}, \epsilon^n_k)$ if the price has increased, and
\item $q^n_{\tau_k}= \tilde{R}_k^n= g(q^n_{\tau_k-}, \tilde{\epsilon}^n_k)$ if the price has decreased,
\end{itemize}
where $(\epsilon^n_k, k \geq 1)$ is an IID sequence with
distribution $f_n$, and $(\tilde{\epsilon}^n_k, k \geq 1)$ is an IID
sequence with distribution $\tilde{f}_n$. Note that this
specification includes Examples \ref{example.IID} and
\ref{example.pegged} as special cases.

We make the following assumptions, which allow for an analytical study of the heavy traffic limit and
  are  sufficiently general to accommodate high frequency data sets of trades and quotes  such as the ones described in Section \ref{sec.empirique}:
\begin{assumption}\label{assumption.G}
 $(T_{i}^{n,a},T^{n,b}_{i})_{i \geq 1}$ is a stationary array of positive random variables
 whose common distribution has a continuous density and  satisfies
 $$ \lim_{n \to \infty} \dfrac{T^{n,a}_{1} + T^{n,a}_{2} + ... + T^{n,a}_{n}}{n} = \dfrac{1}{\lambda^{a}} <\infty, \qquad
 \lim_{n \to \infty} \dfrac{T^{n,b}_{1} + T^{n,b}_{2} + ... + T^{n,b}_{n}}{n} = \dfrac{1}{\lambda^{b}}<\infty. $$
\end{assumption}
$  \lambda^{a}  $ (resp. $\lambda^{b} $) represents the arrival rate of orders at the ask (resp. the bid).
\begin{assumption}\label{assumption.H}   $(V^{n,a}_{i}, V^{n,b}_{i})_{i \geq 1}$ is a  stationary, uniformly mixing array of random variables satisfying
\begin{equation}
 \sqrt{n} \mathbb{E}[V_{1}^{n,a}] \mathop{\rightarrow}^{n \rightarrow \infty} \overline{V^{a}}, \qquad \sqrt{n} \mathbb{E}[V_{1}^{n,b}] \mathop{\rightarrow}^{n \rightarrow \infty} \overline{V^{b}},
\end{equation}
$$\lim_{n \rightarrow \infty} \mathbb{E}[(V_{i}^{n,a}-\overline{V^{a}})^{2}] + 2 \sum_{i=2}^{\infty} {\rm cov}(V_{1}^{n,a},V_{i}^{n,a}) = v_{a}^{2}<\infty, \qquad{\rm and}$$
$$\lim_{n \rightarrow \infty}  \mathbb{E}[(V_{i}^{n,b}-\overline{V^{b}})^{2}] + 2 \sum_{i=2}^{\infty} {\rm cov}(V_{1}^{n,b},V_{i}^{n,b}) = v_{b}^{2}<\infty. $$
\end{assumption}
The assumption of uniform mixing  \cite[Ch. 4]{Billingsley} implies that the partial sums of order sizes verify a central limit theorem, but allows for various types of serial dependence in order sizes.
The scaling assumptions  on the first two moments corresponds to the properties of the empirical data discussed in Section  \ref{sec.fluidornotfluid}.
Under Assumption \ref{assumption.H}, one can define
\begin{equation}\rho:= \lim_{n\to\infty}\dfrac{1}{v_{a} v_{b}}\left( 2 \max( \lambda^{a},\lambda^{b}) {\rm cov}(V_{1}^{n,a},V_{1}^{n,b}) + 2  \sum_{i=1}^{\infty} \lambda^{a} {\rm cov}(V_{1}^{n,a},V_{i}^{n,b}) + \lambda^{b} {\rm cov}(V_{1}^{n,b},V_{i}^{n,a}) \right).\label{eq.rho} \end{equation}
$\rho \in (-1,1)$ may be interpreted as a measure of `correlation' between event sizes at the bid and event sizes at the ask.

These assumptions hold for the examples of Section \ref{sec.examples}. In the case of the Hawkes model, Assumption \ref{assumption.G} was shown to hold in \cite{bacry10}.
Also,  these assumptions are quite plausible for   high frequnecy  quotes  for liquid US stocks since, as argued in Section \ref{sec.empirique}:
\begin{itemize}
\item The tail index of order sizes is larger than two, so the sequences $(V_{i}^{b}, i \geq 1)$ and $(V_{i}^{a}, i \geq 1)$ have a finite second moment.
\item The sequence of order sizes is  uncorrelated i.e. has statistically insignificant autocorrelation. Therefore the sum of autocorrelations of order sizes is finite (zero, in fact).
\item The sequence of inter-event durations has a finite empirical mean and is not autocorrelated.
\end{itemize}
These empirical observations support the plausibility of  Assumptions \ref{assumption.G} and \ref{assumption.H} for the data sets examined.

\begin{center}
\begin{figure}[h!]
\begin{tabular}{cc}
   \includegraphics[width=8cm]{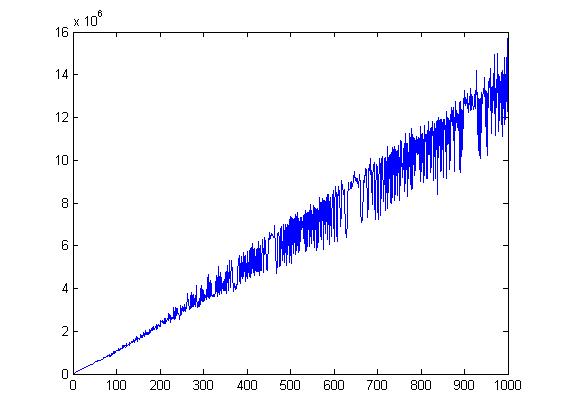} &
   \includegraphics[width=8cm]{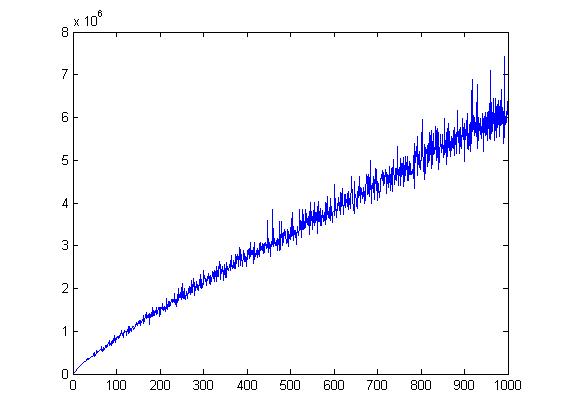} \\
\end{tabular}
\caption{Variance of batch sizes of $n$ orders, for General Electric shares, on June 26th, 2008. Left: ask side. Right: bid side.}\label{varsize.graph}
\end{figure}
\end{center}

Assumption \ref{assumption.H} has an intuitive interpretation: if orders are grouped in batches of $n$ orders, then Assumption \ref{assumption.H} amounts to stating that  the variance of  batch sizes should scale linear with $n$. This assumption can be checked empirically, using a variance ratio test for example: Figure \ref{varsize.graph} shows that this linear relation is indeed verifies for the data sets examined in Section \ref{sec.empirique}.
The following  scaling assumption  states that, when grouping orders in batches of $n$ orders, a good  proportion of batches should have a size  $O(\sqrt{n})$ (otherwise their impact will vanish in the  limit when $n$ becomes large):
\begin{assumption}\label{assumption.F}
There exist  probability distributions $F,\tilde{F}$ on the interior $(0,\infty)\times(0,\infty)$ of the positive orthant, such that
\begin{equation*}
n f_{n}(\sqrt{n}\ .)  \mathop{\Rightarrow}^{n\to\infty} F\quad{\rm and}\quad n \tilde{f}_{n}(\sqrt{n}\ .)  \mathop{\Rightarrow}^{n\to\infty}  \tilde{F}.
\end{equation*}
\end{assumption}
\begin{assumption}\label{assumption.g}
$g\in C^2(\mathbb{R}_+^2\times\mathbb{R}_+^2,]0,\infty[^2)$ and
\begin{equation*}
\exists \alpha>0, \forall (x,y)\in
\mathbb{R}_+^2\times\mathbb{R}_+^2,\qquad \|g(x,y)\|\geq \ \alpha
\min(y_1,y_2).
\end{equation*}
\end{assumption}
Finally, we add the following  condition for the initial value of
the queue sizes:
\begin{equation}\label{eq.initialcondition}
 \left( \frac{q^{n,b}_{0}}{\sqrt{n}}, \frac{q^{n,a}_{0}}{\sqrt{n}}\mathop{\to}^{n\to\infty} (x_0,y_0)\neq (0,0). \right)
\end{equation}
The following theorem, whose proof is given in the Appendix,
describes the joint dynamics of the bid and ask queues in this heavy
traffic limit:
\begin{theorem}[Heavy traffic limit]\label{HeavyTraffic.thm}
Under Assumptions \ref{assumption.F},  \ref{assumption.G}, \ref{assumption.H} and \ref{assumption.g},
the rescaled process
 $$ \left(Q^{n}_t, t \geq 0\right) = \left( \frac{q^{n}_{nt}}{\sqrt{n}}, t \geq 0 \right)$$
converges weakly, on the Skorokhod space $(D(]0,\infty[,\mathbb{R}_+^2),,J_{1})$,
$$Q^{n} \quad\mathop{\Rightarrow}^{n \rightarrow \infty}  Q  $$
 to a Markov process
   $(Q_t)_{t\geq 0}$    with values in $\mathbb{R}_{+}^{2}-\{(0,0)\}$ and infinitesimal generator
$\mathcal{G}$ given  by
\begin{eqnarray} \label{Generator}
\mathcal{G}h(x,y) =  \lambda_{a} \overline{V^a} \dfrac{\partial h}{\partial x} +  \lambda_{b} \overline{V^{b}} \dfrac{\partial h}{\partial y} + \dfrac{ \lambda_{a} v_{a}^{2}}{2} \dfrac{\partial^{2} h}{\partial x^{2}} + \dfrac{ \lambda_{b} v_{b}^{2}}{2} \dfrac{\partial^{2} h}{\partial y^{2}} + 2 \rho \sqrt{\lambda_{a} \lambda_{b}} v_{a} v_{b} \dfrac{\partial^{2} h}{\partial x \partial y} ,\\
 \mathcal{G}h(x,0) =\int_{\mathbb{R}_+^2} \mathcal{G} h( g((x,0),(u,v)) ) F(du,dv),  \quad \mathcal{G}h(0,y) = \int_{\mathbb{R}_+^2} \mathcal{G} h( g((0,y),(u,v)) ) \tilde{F}(du,dv),    \label{eq.boundary}\end{eqnarray}
for $x>0,y>0$, whose  domain is the set
  $ dom(\mathcal{G}) $ of functions $ h \in {C}^{2}(]0,\infty[\times]0,\infty[,\mathbb{R})\cap {C}^{0}(\mathbb{R}_{+}^{2},\mathbb{R})$
  verifying the boundary conditions
\begin{eqnarray*}
\forall x>0,\qquad h(x,0) =  \int_{\mathbb{R}_{+}^2} h(g((x,0),(u,v))) F(du,dv)=0, \\
\forall y>0,\qquad
h(0,y) = \int_{\mathbb{R}_{+}^2} h(g((0,y),(u,v))) \tilde{F}(du,dv)=0.
\end{eqnarray*}
 \end{theorem}
\proof{Proof.} We outline here the main steps of the proof. The technical details are given in the Appendix.
Define the  counting processes
\begin{equation}
 N^{a,n}_{t} = \sup \lbrace k\geq 0, \  T^{a,n}_{1} +  ... + T_{k}^{a,n} \leq t \rbrace  \ \ \ and \ \  N^{b,n}_{t} = \sup \lbrace k\geq 0, \  T^{b,n}_{1} +  ... + T_{k}^{b,n} \leq t \rbrace\label{eq.NaNb}
\end{equation}
which correspond to the number of events at the ask (resp. the bid), and the net order flow
\begin{equation*}
X^n_t=\left(\sum_{i=1}^{ N^{b,n}_{nt}}\dfrac{ V_{i}^{b,n}}{\sqrt{n}},
\sum_{i=1}^{ N^{a,n}_{nt}} \dfrac{V_{i}^{a,n}}{\sqrt{n}}\right)
\end{equation*}
Then, as shown in Proposition \ref{Diffusion-thm} (see Appendix), $X^n$ converges in distribution on $(D([0,\infty[,\mathbb{R}^2), J_1)$ to a two-dimensional Brownian motion with drift
\begin{equation*}
(X^n_t)_{t
\geq 0} \mathop{\Rightarrow}^{n \rightarrow \infty}  \left(Z_{t} + t(\lambda^{b} \overline{V^{b}},\lambda^{a}\overline{V^{a}}) \right)_{t
\geq 0}
\end{equation*}
 where $Z$ is a planar Brownian motion with covariance matrix
\begin{equation*}   \begin{pmatrix}
\lambda^{b} v_{b}^{2} & \rho  \sqrt{\lambda^{a} \lambda^{b}} v_{a} v_{b} \\
\rho  \sqrt{\lambda^{a} \lambda^{b}} v_{a} v_{b} & \lambda^{a}
v_{a}^{2}
\end{pmatrix}.
\end{equation*}
Under assumption \ref{assumption.F}, using the Skorokhod representation theorem, there exist  IID sequences
$((\epsilon^n_k,n\geq 1),(\tilde{\epsilon}^n_k,n\geq 1),\epsilon_k, \tilde{\epsilon}_k)_{k\geq 1}$
and a copy  $X$ of the process
$$\left((x_0,y_0)+ Z_{t} + t(\lambda^{b} \overline{V^{b}},\lambda^{a}\overline{V^{a}}) \right)_{t
\geq 0}$$
 on some probability space $(\Omega_0,{\cal B},\mathbb{Q})$ such that
$\epsilon^n_k\sim f_n, \tilde{\epsilon}^n_k\sim \tilde{f}_n, \epsilon_k\sim F, \tilde{\epsilon}_k\sim \tilde{F}$ and
$$\mathbb{Q}\left(X^n\mathop{\to}^{n\to\infty} X\quad ;\forall k\geq 1, \frac{\epsilon^n_k}{\sqrt{n}}\mathop{\to}^{n\to\infty} \epsilon_k,\quad  \frac{\tilde{\epsilon}^n_k}{\sqrt{n}}\mathop{\to}^{n\to\infty} \tilde{\epsilon_k}\right)=1.$$
Using the notations of Appendix \ref{sec.psicontinuity}, denote by
\begin{itemize}\item $\tau^n_1=\tau(X^n)$ the first exit time of
$X^n$
from the interior $]0,\infty[\times]0,\infty[ $ of the orthant and
\item  $\tau^n_k$   the first exit time of
$\Psi_k(X^n,Q^n_{\tau^n_1},...,Q^n_{\tau^n_{k-1}})$
from  $]0,\infty[\times]0,\infty[ $.
\end{itemize}
We can now construct the process $Q$ by an induction procedure.
Let $\tau_1=\tau(X)$ be the first exit time of $X$ from the orthant. By definition of $Q$, $Q_t=X_t$ for $t< \tau_1$
and, by continuity of the
first-passage time map and the last-evaluation map at a first passage time \cite[Sec. 13.6.3]{Whitt},
$$(\tau^n_1,Q^n_{\tau^n_1-})\mathop{\to}^{n\to\infty} (\tau_1,Q_{\tau_1-})\ \mathbb{Q}-a.s.$$
By definition of $Q$, we have
$$ Q_{\tau_1}=g(X_{\tau_1-},\epsilon_1)
1_{X_{\tau_1}.(0,1)\leq 0}+g(X_{\tau_1-},\tilde{\epsilon}_k) 1_{X_{\tau_1}.(1,0)\leq 0}.$$
$X$ is a continuous process and  the probability thats its path crosses the origin is zero, so
by Lemma \ref{lemma.continuity}, $X$ lies with probability 1 in the continuity set of
 the map $G:\omega\to 1_{\omega_{\tau(\omega)}.(0,1)\leq 0}$. So
using the continuity of $g(.,.)$, we can apply the continuous mapping theorem \cite[Theorem 5.1]{Billingsley},
to conclude that $$ Q^n_{\tau^n_1}  \mathop{\to}^{n\to\infty} Q_{\tau_1}  \quad \mathbb{Q}-a.s.$$
Let us now show the induction step: assume that we  have defined $Q$ on $[0,\tau_{k-1}]$ and shown that
$$(\tau^n_1,..,\tau^n_{k-1},Q^n_{\tau^n_1},...,Q^n_{\tau^n_{k-1}})\mathop{\to}^{n\to\infty} (\tau_1,..,\tau_{k-1},Q_{\tau_1},...,Q_{\tau_{k-1}}) \quad \mathbb{Q}-a.s.$$
Since $\mathbb{Q}( (0,0)\notin \Psi_k(X,Q_{\tau_1},...,Q_{\tau_{k-1}})( [0,\infty))\  )=1$,
Lemma \ref{continuitypsik} implies that
 $(X,Q_{\tau_1},...,Q_{\tau_{k-1}})$ lies with probability 1 in the continuity set of $\Psi_k$, so by
 the continuous mapping theorem
 $$
 \Psi_k(X^n,Q^n_{\tau^n_1},...,Q^n_{\tau^n_{k-1}}) \mathop{\to}^{n\to\infty} \Psi_k(X,Q_{\tau_1},...,Q_{\tau_{k-1}})\quad \mathbb{Q}-a.s.
 $$
Define now $\tau_k$ as the   first exit time of $\Psi_k(X
,Q_{\tau_1},...,Q^n_{\tau_{k-1}})$ from
$]0,\infty[\times]0,\infty[$. As before, by continuity of the
first-passage time map and the last-value map at a first passage
time \cite[Sec. 13.6.3]{Whitt},
$$(\tau^n_k,Q^n_{\tau^n_k-})\mathop{\to}^{n\to\infty} (\tau_k,Q_{\tau_k-})\quad \mathbb{Q}-a.s.$$
We can now extend the definition of $Q$ to $[0,\tau_{k}]$ by setting
$$ Q_t= \Psi_k(X ,Q_{\tau_1},...,Q_{\tau_{k}})(t)\qquad {\rm for}\quad t<\tau_k , \qquad {\rm and} $$
$$ Q_{\tau_k}=g(Q_{\tau_k-},\epsilon_k)
1_{\Psi_k(X ,Q_{\tau_1},...,Q_{\tau_{k-1}}).(0,1)\leq 0}+g(Q_{\tau_k-},\tilde{\epsilon}_k) 1_{\Psi_k(X ,Q_{\tau_1},...,Q_{\tau_{k-1}}).(1,0)\leq 0} $$
As above,  using the continuity properties of $\Psi_k$ from Lemma \ref{continuitypsik} we conclude that $Q^n_{\tau^n_k}\to Q_{\tau_k}$ a.s.
So finally, we have shown that
$$\forall k\geq 1, \qquad (\tau^n_1,..,\tau^n_{k},Q^n_{\tau^n_1},...,Q^n_{\tau^n_{k}})\mathop{\to}^{n\to\infty} (\tau_1,..,\tau_{k},Q_{\tau_1},...,Q_{\tau_{k}})\qquad  \mathbb{Q}-a.s.$$
We can now   construct the sequences $R,\tilde{R}$ by setting
\begin{itemize}
  \item $R_k=Q_{\tau_{k}}$ if $\Psi_k(X ,Q_{\tau_1},...,Q_{\tau_{k-1}})(\tau_{k}-).(0,1) \leq
  0$,
  \item $\tilde{R}_k=Q_{\tau_{k}}$ if $\Psi_k(X ,Q_{\tau_1},...,Q_{\tau_{k-1}})(\tau_{k}-).(1,0) \leq
  0$.
\end{itemize}
Then $Q=\Psi(X,R,\tilde{R})$ where  $\Psi$ is the map defined in
Definition \ref{def.regularization}. Let us now show that
$(X,R,\tilde{R})$ lies  with probability 1 in the $J_1-$continuity
set of $\Psi$, in order to apply the continuous mapping theorem.
 $X$ is a continuous process
whose paths lie in $C^0([0,\infty),\mathbb{R}^2-\{(0,0)\})$ almost
surely. Since $F$ and $\tilde{F}$ have zero mass on the axes, with
probability 1 the sequences $(\epsilon_k)_{k\geq 1},
(\tilde{\epsilon}_k)_{k\geq 1}$ do not have any accumulation point
on the axes. Assumption \ref{assumption.g} then implies that the
sequences $(R_k)_{k\geq 1}, (\tilde{R}_k)_{k\geq 1}$ do not have any
accumulation point on the axes.
 From the
definition of $\Psi$ (Definition \ref{def.regularization}), $Q$
jumps at each hitting time of the axes and, in between two jumps,
its increments follow those of the planar Brownian motion $X$. Since
$F, \tilde{F}$ have no mass at the origin and   planar Brownian paths
have a zero probability of hitting isolated points,  with
probability 1 the graph of $Q=\Psi(X,R,\tilde{R})$ does not hit the
origin :
\begin{equation}
\mathbb{Q}\left(\ (0,0)\notin \Psi(X,R,\tilde{R})( [0,\infty)\ )\ \right)=1.
\end{equation}
So the triplet $(X,R,\tilde{R})$ satisfies the conditions of Theorem \ref{prop.psicontinuity} almost-surely i.e.
$\Psi$ is continuous at $(X,R,\tilde{R})$ with probability 1.
We can therefore apply the continuous mapping theorem \cite[Theorem 5.1]{Billingsley} and conclude that
\begin{equation*}
Q^{n}=(X^n,R^n,\tilde{R}^n) \mathop{\Rightarrow}^{n\to\infty} Q=\Psi(X,R,\tilde{R}).
\end{equation*}
The process $Q=\Psi(X,R,\tilde{R})$ can be explicitly
construction from the planar Brownian motion $X$ and the sequences $R,\tilde{R}$: $Q$ follows the
increments of $X$ and is reinitialized to  $R_n$ or $\tilde{R}_n$ at
each hitting time of the axes. Lemma \ref{lemma.generator} in
Appendix \ref{appendix.generator} uses this description to show that
$Q$ is a Markov process whose generator is given by
\eqref{Generator}- \eqref{eq.boundary}.
\endproof

\begin{remark}[L\'evy process limits]
The diffusion approximation inside the orthant fails when order sizes do not have a finite second moment. For example, if
the sequence $(V_{i}^{a},V_{i}^{b})$ is regularly varying with tail exponent $\alpha \in (0,2)$ (see \cite{resnick} for definitions),
the heavy-traffic approximation $Q$ is a  pure-jump process in the positive orthant, constructed by applying the map $\Psi$ to a two-dimensional
 $\alpha$-stable  L\'evy process $L$:
$$ Q= \Psi(L,R,\tilde{R}),$$
i.e. by re-initializing it according to \eqref{eq.generalRn}  at each attempted exit from the positive orthant.
We do not further develop this case here, but it may be of interest for the study of illiquid limit order markets, or those where order flow is dominated by large investors.
\end{remark}

\subsection{Jump-diffusion approximation for order book
dynamics}\label{sec.Markovapproximation}

 Theorem
\ref{HeavyTraffic.thm} implies that, when examined over time scales
much larger than the interval between order book events, the queue sizes
$q^{b}$ and $q^{a}$ are well described by a Markovian jump-diffusion
process  $(Q_t)_{t\geq 0}$  in the positive orthant $\mathbb{R}_+^2$ which
  behaves like a a planar Brownian motion with drift  vector \begin{equation}    \label{eq.driftvector}
(\lambda^{b} \overline{V^{b}},\lambda^{a}\overline{V^{a}})\end{equation}  and covariance matrix
\begin{equation}    \label{eq.matrix}
\begin{pmatrix}
\lambda^{b} v_{b}^{2} & \rho  \sqrt{\lambda^{a} \lambda^{b}} v_{a} v_{b} \\
\rho  \sqrt{\lambda^{a} \lambda^{b}} v_{a} v_{b} & \lambda^{a}
v_{a}^{2}
\end{pmatrix}.
\end{equation}
 in the interior $]0,\infty[^2$ of the orthant
and, at each hitting time $\tau_k$ of the axes, jumps to a new position
\begin{itemize}
\item $Q_{\tau_k}= R_k = g( Q_{\tau_k-}, \epsilon_k)$ if it hits the horizontal axis,
\item $Q_{\tau_k}= \tilde{R}_k= g( Q_{\tau_k-}, \tilde{\epsilon}_k)$ if it hits the vertical axis,
\end{itemize}
where the $\epsilon_k$ are IID  with
distribution $F$ and the $\tilde{\epsilon}_k$ are IID
 with distribution $\tilde{F}$.
We note that similar processes in the orthant were studied by \cite{baccelli87} with  queueing applications in mind, but not in the context of heavy traffic limits.

This
process is analytically and computationally tractable  and
allows various quantities related to intraday price behavior to be
computed (see next section).

 If $\gamma_{0} = (\mathbb{E}[T_{1}^{a}] + \mathbb{E}[T_{1}^{b}])/2$ is the average time between order book events,  ($\gamma_{0} \leq $ 100 milliseconds), and $\gamma_1\gg \gamma_0$ (typically, $\gamma_1\sim$ 10-100 seconds) then Theorem
\ref{HeavyTraffic.thm} leads to an approximation for the distributional properties of the queue dynamics in terms of $Q_t$:
$$  q_t \simeq^{d}  \sqrt{N}\quad Q_{t/N}\qquad {\rm where}\qquad N=\frac{\gamma_1}{\gamma_0}$$
So, under Assumptions \ref{assumption.F}, \ref{assumption.G}, \ref{assumption.H} and
\ref{assumption.g}  the order book process $(q_{t}^{b},q_{t}^{a})_{t
\geq 0}$ at the time scale $\gamma_{1}$ can be approximated by a
Markov process  which
\begin{itemize}
\item behaves like a two-dimensional Brownian motion with drift $(\mu_{b},\mu_{a})$ and covariance matrix $\Lambda$ on $\lbrace x > 0 \rbrace \cap \lbrace y > 0 \rbrace$ with
\begin{equation} \label{eq.lambda}
\mu_{a} =    \sqrt{N}  \lambda_{a} \overline{V^a}, \ \qquad \ \ \mu_{b} = \sqrt{N} \lambda_{b}\overline{V^b},
\qquad
\Lambda  = N \begin{pmatrix}
\lambda^{b} v_{b}^{2} & \rho \sqrt{\lambda^{a} \lambda^{b}} v_{a} v_{b} \\
\rho \sqrt{\lambda^{a} \lambda^{b}} v_{a} v_{b} & \lambda^{a} v_{a}^{2}
\end{pmatrix}
\end{equation}
and, at at each hitting time of the axes,
\item jumps to a new value $g( q_{t-}, \sqrt{N} \epsilon_k )$  if  $ q_{t-}^a = 0 $,
\item jumps to a new value $g( q_{t-}, \sqrt{N} \tilde{\epsilon}_k )$  if  $ q_{t-}^b = 0 $,
\end{itemize}
where $\epsilon_k\sim F$, $\tilde{\epsilon}_k\sim \tilde{F}$ are  IID.

This gives a rigorous justification for modeling the queue sizes by a diffusion process at such intraday time scales, as proposed in  \cite{avellaneda2010}.
The parameters involved in this approximation are straightforward to estimate from empirical data: they involve estimating first and second moments of durations and order sizes.

\begin{example}
Set  for instance $\gamma_{1} = 30$ seconds and $\gamma_{0} = (\mathbb{E}[T^{a}] +  \mathbb{E}[T^{b}])/2$.
The following table shows the parameters \eqref{eq.lambda} estimated from high frequency records or order book events for three liquid US stocks.
\begin{table}[h]
\begin{center}

\begin{tabular}{|l|c|c|c|c|c|}
\hline
 & Std deviation of &  Std deviation of &  &  & \\
 & Bid queue &  Ask queue  & $\mu_{b}$ &  $\mu_{a}$& $\rho$\\ \hline

Citigroup & 6256  & 4457  & -1033  & -2467 & 0.07  \\ \hline

General Electric  &2156 & 2928  & -334  & -1291 & 0.03\\ \hline

General Motors  & 578 & 399  & +78  & -96  & - 0.04 \\ \hline
\end{tabular}\vspace{0.5cm}
\caption{Parameters for the heavy-traffic approximation of bid / ask queues over a 30-second time scale. The unit is a number of orders per period of 30 seconds.}
\end{center}\label{volatility.table}
\end{table}

In particular we observe that the order of magnitude of the standard deviation of queue lengths is an order of magnitude larger  than their expected change.
\end{example}

\begin{example}  Theorem \ref{HeavyTraffic.thm} may also be used to derive jump-diffusion approximations for the limit order book in theoretical models such as the ones presented in Section  \ref{sec.examples}.
Let us illustrate this in the case of the heterogeneous trader model of Section \ref{sec.ordersplit}.

Let $(T_{i}, i \geq 1)$ the sequence of duration between consecutive
orders. We assume that this sequence is a sequence of stationary
random variables with $\mathbb{E}[T_{1}] < \infty$. We also assume
that every trader has an equal chance of being a buyer or a seller
and that the type of trader (buyer or seller) is independent from
the past:

\begin{equation*}
\mathbb{P}[i-\text{th trader is a buyer}] =  \mathbb{P}[i-\text{th trader is a seller}] = \frac{1}{2}
\end{equation*}

Finally the sequence of number of orders $(V_{i}, i \geq 1)$ is a stationary sequence of orders traded by the $i$-th trader with the property that $\mathbb{E}[V_{1}^{2}] < \infty$.

This order flow given by $(T_{i}, i \geq 1)$, $(V_{i}, i \geq 1)$, and the sequence of type (buyers or sellers, using limit orders, market orders or both) generates a sequence of durations $(T_{i}^{a}, i \geq 1)$, $(T_{i}^{b}; i \geq 1)$ and order sizes $(V_{i}^{a}, i \geq 1)$ and $(V_{i}^{b}, i \geq 1)$ which satisfy assumptions \ref{assumption.G} and \ref{assumption.H}.

The sequence of durations $(T_{i}^{a}, i \geq 1)$ and $(T_{i}^{b}, i \geq 1)$ are two stationary sequences of random variables with finite mean:

\begin{equation*}
\forall i \geq 0, \ \ T_{i} = T_{i}^{a} = T_{i}^{b}. \ \ \text{therefore}  \ \ \mathbb{E}[T_{i}] = \mathbb{E}[T_{i}^{a}] = \mathbb{E}[T_{i}^{b}] < \infty.
\end{equation*}

The sequence of order sizes $((V_{i}^{b},V_{i}^{a}), i \geq 1)$  is a sequences of IID random variables with
\begin{equation}
\mathbb{P}[(V_{i}^{b},V_{i}^{a}) = (V_{i},0)] = \mathbb{P}[(V_{i}^{b},V_{i}^{a}) = (0,V_{i})] = \frac{m}{2},
\end{equation}
\begin{equation}
\mathbb{P}[(V_{i}^{b},V_{i}^{a}) =( -V_{i},0)] = \mathbb{P}[(V_{i}^{b},V_{i}^{a}) =( 0,-V_{i})] = \frac{l}{2},
\end{equation}
\begin{equation}
\mathbb{P}[(V_{i}^{b},V_{i}^{a}) = (\gamma V_{i}, -(1- \gamma)V_{i})]= \mathbb{P}[(V_{i}^{b},V_{i}^{a}) = (-(1-\gamma) V_{i}, \gamma V_{i})] = \frac{1 - l - m}{2}.
\end{equation}
Theorem \ref{HeavyTraffic.thm} then shows that $(Q^{b},Q^{a})$  is a Markov process which behaves like a two-dimensional Brownian motion with drift $(\mu_{b},\mu_{a})$ and covariance matrix $\Lambda$ inside the positive orthant $\lbrace x > 0 \rbrace \cap \lbrace y > 0 \rbrace$  where:
\begin{equation}
\mu_{b} = \mu_{a} = \frac{\overline{V}}{2 \mathbb{E}[T_{1}]} \left( 2m + 2 \gamma(1 - l - m) -1 \right), \ \
\Lambda  = v^{2} \begin{pmatrix}
 1 & \rho   \\
\rho  & 1
\end{pmatrix}, \ \ where
\end{equation}
\begin{equation}
v^{2} = \dfrac{\mathbb{E}[T_{1}] \mathbb{E}[V_{1}^{2}]}{4} \left( m + l + \frac{\gamma^{2} + (1 - \gamma)^{2}}{2}(1 - l - m) \right) \ \ and \ \ \rho = - \dfrac{(1 - l - m)^{2} \gamma (1 - \gamma)}{1 + (1 - l - m)(\gamma^{2} - \gamma - 1/2)}  < 0.
\end{equation}
Figure \ref{correlgamma} displays the value of the correlation $\rho$ in different scenarios as a function of $\gamma$ and the proportion $1-(l+m)$ of  traders submitting orders of both types.
\begin{center}
\begin{figure}[tbh]
 \centering
   \includegraphics[width=10cm]{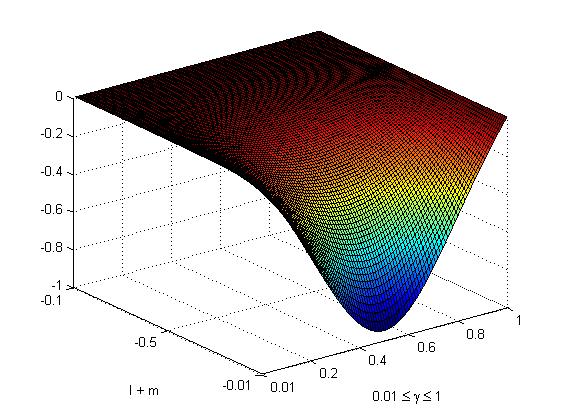}
\caption{Correlation $\rho$ between bid and ask queue sizes for different scenario. 1 - $(l + m)$ represents the proportion of traders using both market and limit orders, $\gamma$ the proportion of limit orders and $(1- \gamma)$ the proportion of market orders.}
 \label{correlgamma}\end{figure}
\end{center}
\end{example}
\clearpage

\section{Price dynamics}\label{sec.computations}

\subsection{Price dynamics in the heavy traffic limit}

Denote by $(s^{n}_{t}, t \geq 0)$ the (bid) price process corresponding to the limit order book process $(q^n_t)_{t \geq 0}$.  As explained in Section \ref{sec.model}, $s^{n}$ is a piecewise constant stochastic process which\begin{itemize}\item
increases by one tick at each event $(t^{a,n}_i,V^{a,n}_i)$ at the ask for which $$q^{a,n}(t^{a,n}_i)+\frac{V^{a,n}_i}{\sqrt{n}} \leq 0 $$ \item decreases by one tick at each event $(t^{b,n}_i,V^{b,n}_i)$ at the bid for which $$q^{b,n}(t^{b,n}_i)+\frac{V^{b,n}_i}{\sqrt{n}} \leq 0 .$$\end{itemize}
Due to the complex dependence structure in the sequence of order durations and sizes, properties of the process $s^n$ are not easy to  study, even in simple models such as those given in Section  \ref{sec.examples}. The following result shows that the price process converges to a simpler process in the heavy traffic limit, which is entirely characterized by hitting times of the two dimensional Markov $Q$:
\begin{proposition}\label{prop.pricedynamics} Under the assumptions of Theorem  \ref{HeavyTraffic.thm},
\begin{equation*}
(s^{n}_{nt},  t\geq 0) \mathop{\Rightarrow}^{n\to\infty} S, \quad \text{on} \quad ( D([0,\infty[,\mathbb{R}),M_{1}), \qquad  {\rm where}
\end{equation*}
\begin{equation} S_t = \sum_{0 \leq s\leq t } \mathbf{1}_{Q^a_s=0}  - \sum_{0 \leq s\leq t } \mathbf{1}_{Q^b_s=0}
\end{equation}   is a piecewise constant cadlag process which
\begin{itemize} \item
increases by one tick every time the process  $Q$ hits the horizontal axis
$\lbrace y = 0 \rbrace$ and \item decreases by one tick every time $Q$ hits the vertical axis $\lbrace x = 0 \rbrace$.\end{itemize}
\end{proposition}
\proof{Proof.}  We refer the reader to \cite{Whitt} or \cite{whitt80} for a description of the $M_1$ topology.
The price process $s^n$ (rescaled in time) can be expressed as
\begin{eqnarray}
  s^n_{nt} & = & \sum_{\tau_{k}^n \leq t} 1_{\Psi_k(X^n,Q^n_{\tau^n_1},...,Q^n_{\tau^n_{k-1}})(\tau^n_k).(0,1) \leq 0 }
   - 1_{\Psi_k(X^n,Q^n_{\tau^n_1},...,Q^n_{\tau^n_{k-1}})(\tau^n_k).(1,0) \leq 0 }. \nonumber
\end{eqnarray}
where $\tau_k^n, Q^n_{\tau^n_{k}}$ are defined in the proof of
Theorem \ref{HeavyTraffic.thm}. There we showed that
$$\forall k\geq 1, \quad (X^n,\tau^n_1,..,\tau^n_{k},Q^n_{\tau^n_1},...,Q^n_{\tau^n_{k}})\mathop{ \Rightarrow}^{n\to\infty} (X,\tau_1,..,\tau_{k},Q_{\tau_1},...,Q_{\tau_{k}}).$$
As shown in the proof of Theorem \ref{HeavyTraffic.thm},
$(X,Q_{\tau_1},...,Q_{\tau_{k},...})$ lies, with probability $1$, in
the set of continuity points of $\Psi_k$ for each $k\geq 1$ so
$$ \Psi_k(X^n,Q^n_{\tau^n_1},...,Q^n_{\tau^n_{k-1}}) \mathop{ \Rightarrow}^{n\to\infty} \Psi_k(X,Q_{\tau_1},...,Q_{\tau_{k-1}}).$$
Applying Proposition \ref{lemma.continuity} (see Appendix) and the continuous mapping theorem \cite[Theorem 5.1]{Billingsley} then shows
that
$$ 1_{\Psi_k(X^n,Q^n_{\tau^n_1},...,Q^n_{\tau^n_{k-1}})(\tau^n_k).(0,1) \leq 0 }\mathop{ \Rightarrow}^{n\to\infty}
1_{\Psi_k(X,Q_{\tau_1},...,Q_{\tau_{k-1}})(\tau_k).(0,1) \leq 0 }
$$

The sequences of processes $\sum_{\tau_{k}^n \leq t} 1_{\Psi_k(X^n,Q^n_{\tau^n_1},...,Q^n_{\tau^n_{k-1}})(\tau^n_k).(0,1) \leq 0 }$ and $\sum_{\tau_{k}^n \leq t} 1_{\Psi_k(X^n,Q^n_{\tau^n_1},...,Q^n_{\tau^n_{k-1}})(\tau^n_k).(1,0) \leq 0 }$ belong to $D_{\uparrow}([0,\infty[,\mathbb{R}_+)$, the set of increasing cadlag trajectories. The convergence on the $M_{1}$ topology for sequences in $D_{\uparrow}$ is equivalent reduces to the convergence on a dense subset including zeros. Therefore,

\begin{equation*}
\sum_{\tau_{k}^n \leq t} 1_{\Psi_k(X^n,Q^n_{\tau^n_1},...,Q^n_{\tau^n_{k-1}})(\tau^n_k).(0,1) \leq 0 } \Rightarrow \sum_{\tau_{k} \leq t} 1_{\Psi_k(X,Q_{\tau_1},...,Q_{\tau_{k-1}})(\tau_k).(0,1) \leq 0 }, \quad \text{and}
\end{equation*}
\begin{equation*}
\sum_{\tau_{k}^n \leq t} 1_{\Psi_k(X^n,Q^n_{\tau^n_1},...,Q^n_{\tau^n_{k-1}})(\tau^n_k).(1,0) \leq 0 } \Rightarrow \sum_{\tau_{k} \leq t} 1_{\Psi_k(X,Q_{\tau_1},...,Q_{\tau_{k-1}})(\tau_k).(1,0) \leq 0}.
\end{equation*}

On the other hand, since the set of discontinuities of $\sum_{\tau_{k} \leq t} 1_{\Psi_k(X,Q_{\tau_1},...,Q_{\tau_{k-1}})(\tau_k).(0,1) \leq 0 }$ and $\sum_{\tau_{k} \leq t} 1_{\Psi_k(X,Q_{\tau_1},...,Q_{\tau_{k-1}})(\tau_k).(1,0) \leq 0}$ have an intersection which is almost surely void, one can apply \cite[Theorem 4.1]{whitt80} and \cite[Theorem 12.7.1]{Whitt} and
\begin{equation*}
(s^{n}_{nt},  t\geq 0) \mathop{\Rightarrow}^{n\to\infty} S. \quad \text{on} \quad (D([0,\infty[,\mathbb{R}_+),M_{1}).
\end{equation*}
\endproof
$S$ is thus the difference between the  occupation time of the $y$ axis and the occupation time of the $x$ axis by the Markov process $Q$.
In particular, this result shows that, in a market where order arrivals are frequent, distributional properties of the price process $s^n$ may be approximated using the distributional properties of the limit $S$. We will now use this result to obtain some  analytical results on the distribution of durations between price changes and the transition probabilities of the price.
\subsection{Duration between price moves} \label{TimeBrownian}

Starting from an initial order book configuration $Q_0=(x,y)$, consider the  hitting times
\begin{itemize}
\item  the next price increase occurs at the first hitting time of the $x$-axis by $Q$:
$$ \tau_a \mathop{=}   \inf\{ t\geq 0, Q^a_t=0 \}$$
\item  the next price decrease occurs at the first hitting time of the $x$-axis by $Q$:
$$ \tau_b \mathop{=}   \inf\{ t\geq 0, \quad Q^b_t=0 \}.$$
\end{itemize}
The  {\it duration}  $\tau$ until the next price changes is then  given by
$$ \tau = \tau_{a} \wedge \tau_{b}, $$
which has the same law as the first exit  time from the positive orthant of a two-dimensional Brownian motion  with drift. Using the results of \cite{Iyengar,Metzler,Zhou2001} we obtain the following
 result which relates the distribution of this {\it duration}  to the state of the order book and the
statistical feature of the order flow process in the case wbalanced order flow where $\overline{V^a} = \overline{V^b} = 0$.
\begin{proposition}[Conditional distribution of duration between price changes]
In a balanced order flow where $\overline{V^a} = \overline{V^b} = 0$  the distribution of the duration $\tau$ until the next price change, conditonal on the current state of the bid and ask queues, is given by
$$ \mathbb{P}[\tau > t| Q_{0}^{b} = x, Q_{0}^{a} = y] = \sqrt{\dfrac{2 U }{ \pi  t}} e^{- \dfrac{U}{4  t}} \sum_{n=0}^{\infty} \dfrac{1}{(2n+1)} \sin \dfrac{(2n+1) \pi \theta_{0}}{\alpha} (I_{(\nu_{n}-1)/2}(\dfrac{U}{4 t}  )  + I_{(\nu_{n}+1)/2}(\dfrac{U}{4 t} )  ), $$
where $\nu_{n} = (2n+1)\pi/\alpha$, $I_{n}$ is the $n$th Bessel function,
$$U= \dfrac{ ( \frac{x}{\lambda_av_a^2})^2 + ( \frac{y}{\lambda_bv_b^2})^2 -2
\rho \frac{xy}{\lambda_a\lambda_b v_a^2v_b^2} }{ (1-\rho)},\qquad {\rm and}$$
\begin{eqnarray} \label{alpha}
\alpha =
\left\{
\begin{array}{lcl}
\displaystyle
\pi + \tan^{-1}(-\dfrac{\sqrt{1-\rho^{2}}}{\rho}) &  & \rho > 0\\
\\
 \dfrac{\pi}{2} &   & \rho = 0\\
 \\
 \tan^{-1}(-\dfrac{\sqrt{1-\rho^{2}}}{\rho}) &  & \rho < 0\\
\end{array}
\right. {\rm and}\qquad
\theta_{0} =
\left\{
\begin{array}{lcl}
\displaystyle
\pi + \tan^{-1}(-\dfrac{y\sqrt{1-\rho^{2}}}{x -\rho y}) &  & x < \rho y\\
\\
 \dfrac{\pi}{2} &   & x = \rho y\\
 \\
 \tan^{-1}(-\dfrac{y\sqrt{1-\rho^{2}}}{x -\rho y}) &  & x > \rho y\\
\end{array}
\right.
\end{eqnarray}
In particular, $\tau$ is regularly varying with tail index $\dfrac{\pi}{2 \alpha}$.
\end{proposition}
\proof{Proof.}
When  $\overline{V^a} = \overline{V^b} = 0$, the process $Q$ behaves like a two-dimensional Brownian motion $Z$ with covariance matrix given by   \eqref{eq.matrix} up to the first hitting time of the axes, so  the distribution of the duration $\tau$ has the same law as  the first exit time of $Z$ from the orthant:
$$ \tau \mathop{=}^d  \inf\{ t\geq 0, Q^a_t=0\quad{\rm or}\  Q^b_t=0 \}$$
Using the results of \cite{Iyengar}, corrected by \cite{Metzler} for the distribution of the first exit time of a two-dimensional Brownian motion  from the orthant we obtain  the result.

A result of \cite{Spitzer} then shows that
$$ \mathbb{E}[\tau^{\beta}| Q_{0}^{b}= x, Q_{0}^{a} = y] = \int_{0}^{\infty} t^{\beta - a} \mathbb{P}[\tau > t| Q_{0}^{b}= x, Q_{0}^{a} = y] dt  < \infty$$
if and only if  $\beta < \pi/2 \alpha$, where $\alpha$ is defined in  \eqref{alpha}. Therefore the tail index of $\tau $ is $\dfrac{\pi}{2 \alpha}$. This result does not depend on the initial state $(x,y)$.
\endproof
\begin{itemize}
\item If $\rho = 0$, the two components of the Brownian Motion are independent and $\tau$ is a regularly-varying random variable with tail index $1$. This random variable does not have a moment of order one.
\item If $\rho < 0$, $\dfrac{\pi}{2 \alpha} >1$ and $\tau$ has a finite moment of order one. In practice, $\rho \approx -0.7$; this means that if $\mu_{a} = 0$ and $\mu_{b} = 0$, the tail index of $\tau$ is around $2$.
\item When $\rho > 1 $, $\dfrac{\pi}{2 \alpha} <1$. The tail of $\tau$ is very heavy; $\tau$ does not have a finite moment of order one.
\end{itemize}
For all high frequency data sets examined,  the estimates for $\mu_{a}, \mu_{b} $ are negative (see Section \ref{sec.Markovapproximation}); the durations then  have finite moments of all orders.

\begin{remark}
Using the results of \cite{Zhou2001} on the first exit time of a two-dimensional Brownian motion with drift, one can generalize the above results to the case where $(\overline{V^{b}},\overline{V^{a}})\neq (0,0)$: we obtain in that case
\begin{equation}
\mathbb{P}[\tau > t| Q_{0}^{b} = x, Q_{0}^{a} = y] = \dfrac{2e^{a_{1}x_{1} + a_{2}x_{2} + a_{t} t -r_{0}^{2}/2t}}{\alpha t}  \sum_{n=1}^{\infty} \sin\left( \dfrac{n \pi \theta_{0}}{\alpha} \right) \int_{0}^{\alpha} \sin \left( \dfrac{n \pi}{\alpha} \right) g_{n}(\theta) d \theta
\end{equation}
where $\theta_{0}, \alpha$ are defined as above, $r_{0} = \sqrt{U}$ and
\begin{equation*}
g_{n}(\theta) = \int_{0}^{\infty} re^{-r^{2}/2t} e^{d_{1}r \sin(\theta - \alpha)- d_{2} r \cos(\theta - \alpha)} I_{n\pi/ \alpha}(\dfrac{r r_{0}}{t})dr,
\end{equation*}
\begin{eqnarray}
d_{1} = \left(a_{1} \sqrt{\lambda_{a}} v_{a} + \rho a_{2} v_{b} \sqrt{\lambda_{b}}\right), \ \ d_{2} =  \left(\rho a_{1} \sqrt{\lambda_{a}} v_{a} +  a_{2} v_{b} \sqrt{\lambda_{b}}\right)\\
a_{1} = -  \dfrac{ \mu_{a} \sqrt{\lambda_{b}} v_{b} + \mu_{b} \rho v_{a} \sqrt{\lambda_{a}}}{(1- \rho^{2}) \sigma^{2}_{a} \lambda_{a} \sqrt{\lambda_{b}} v_{b}}, \qquad
a_{2} = -  \dfrac{ \rho \mu_{a} \sqrt{\lambda_{b}} v_{b} + \mu_{b}  v_{a} \sqrt{\lambda_{a}}}{(1- \rho^{2}) \sigma^{2}_{b} \lambda_{b} \sqrt{\lambda_{a}} v_{a}},
\\
{\rm and}\qquad a_{t} =  \left(a_{1} \dfrac{\lambda_{a} v_{a}^{2}}{2} + a_{2} \dfrac{\lambda_{b} v_{b}^{2}}{2} + 2 \rho a_{1} \sqrt{\lambda_{a} \lambda_{b}} v_{a} a_{2} v_{b} \right) - a_{1} \mu_{a} - a_{2} \mu_{b}.\qquad
\end{eqnarray}

\end{remark}

\subsection{Probability of a price increase} \label{DirichletBrownian}

A useful quantity for short-term prediction of intraday price moves is the
probability $p^{up}(x,y)$   that the price will increase at the
next move given $x$ orders at the bid and $y$ orders at the ask; in our setting this is equal to the probability that the ask queue gets depleted
before the bid queue.

In the heavy traffic limit, this quantity may be represented as the probability that the
two-dimensional process $(Q_t, t \geq 0)$, starting from an initial position $(x,y)$, hits the horizontal axis before
hitting the vertical axis:
$$ p^{up}(x,y) = \mathbb{P}[\tau_a< \tau_b |(Q_{0}^{b},Q_{0}^{a}) = (x,y) ].$$
Since this quantity only involves the process $Q$ up to its first hitting time  of the boundary of the orthant, it may be equivalently computed by replacing $Q$ by a two-dimensional Brownian motion with drift and covariance given by \eqref{eq.driftvector}--\eqref{eq.matrix}.

However, when $\overline{V^a} = \overline{V^b} = 0$,
one has a simple analytical solution which only depends on the size  $x$ of the  bid queue,
the  size $y$ of the ask queue and the correlation
$\rho$ between their increments:
\begin{theorem} Assume $\overline{V^a} + \overline{V^b} \leq  0$.
Then $p^{up}:\mathbb{R}_+^2\to [0,1]$ is the unique bounded solution of the Dirichlet  problem:
\begin{equation}
\dfrac{ \lambda_{a} v_{a}^{2}}{2} \dfrac{\partial^{2} p^{up}}{\partial x^{2}} + \dfrac{ \lambda_{b} v_{b}^{2}}{2} \dfrac{\partial^{2} p^{up}}{\partial y^{2}} + 2\rho \sqrt{\lambda_{a} \lambda_{b}} \sigma^{a} \sigma^{b}  \dfrac{\partial^{2} p^{up}}{\partial x \partial y} +  \lambda_{a} \overline{V^a} \dfrac{\partial p^{up}}{\partial x} +  \lambda_{b} \overline{V^b} \dfrac{\partial p^{up}}{\partial y}= 0\quad{\rm for}\quad{x>0,\quad y>0}\quad\label{eq.Dirichlet}
\end{equation}
with the boundary conditions
\begin{equation}
 \forall x>0,  \quad p^{up}(x,0) = 1 \quad \ { \rm and } \quad \forall y>0, \quad p^{up}(0,y) = 0.\label{eq.pupboundary}
\end{equation}
When $\overline{V^a} = \overline{V^b} = 0$,
 $p_{\rm up}(x,y)$ is given by
\begin{equation}
p_{\rm up}(x,y)=\frac{1}{2}- \dfrac{ \arctan(\sqrt{ \frac{1+\rho}{1-\rho}} \frac{\frac{y}{\sqrt{\lambda_a}v_a}-\frac{x}{\sqrt{\lambda_b}v_b}}{\frac{y}{\sqrt{\lambda_a}v_a}+
\frac{x}{\sqrt{\lambda_b}v_b} } ) } {2\arctan(\sqrt{ \frac{1+\rho}{1-\rho} })}\label{pupW.eq},
\end{equation}
where $\lambda_{a}, \lambda_{b}, v_{a}$ and $v_{b}$ are defined \ref{assumption.G} and \ref{assumption.H}.
\end{theorem}
\proof{Proof.}
Using the results of \cite{yoshida99}, the Dirichlet problem  \eqref{eq.Dirichlet}--\eqref{eq.pupboundary} has a unique positive bounded solution $ u\in  C^2(]0,\infty[^2,\mathbb{R}_+)\cap C^0_b(\mathbb{R}_+^2,\mathbb{R}_+)$.
Application of Ito's formula to $M_t=u(Q^b_t,Q^a_t)$ then shows that the process $M^\tau$  stopped at $\tau$ is a martingale, and conditioning with respect to $(Q^b_0,Q^a_0)=(x,y)$ gives $u(x,y)=p^{\rm up}(x,y)$

Assume now $\overline{V^a} = \overline{V^b} = 0$.
Using a change of variable $x \mapsto x \ \sqrt{\lambda_{b}} v_{b}$ and $y \mapsto y \ \sqrt{\lambda_{a}} v_{a}$,
one only needs to consider  the case where $\sqrt{\lambda_{b}} v_{b} = \sqrt{\lambda_{a}} v_{a}$.

\begin{center}
\begin{figure}[tbh]
\centering
\includegraphics[width=0.7\textwidth]{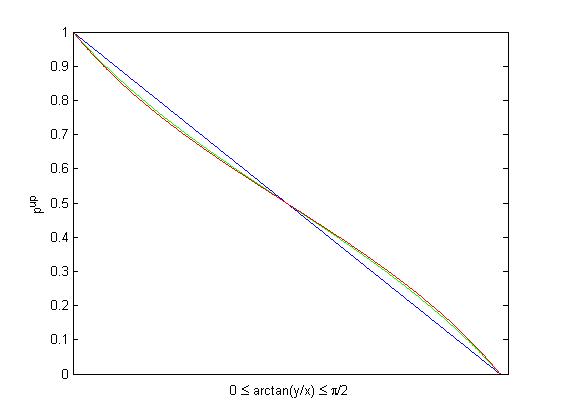}
\caption{$p^{up}$ as a function of the bid-ask imbalance variable $\theta=\arctan(y/x)$ for $\rho = 0$ (blue line), $\rho = -0.7$ (green line) and $\rho = -0.9$ (red line).}\label{graphpup}
\end{figure}
\end{center}

Up to the first hitting time of the axes, $(Q_{t}, t \geq 0 )$ is identical in law to $Q = A B$ where
$$ A =  \begin{pmatrix}
\cos(\beta) & \sin(\beta) \\
\sin ( \beta)& \cos(\beta)
\end{pmatrix}, $$
with $\beta$ satisfying $\rho = \sin(2 \beta)$, $\beta \leq \pi/4$ and $B$ a standard planar Brownian Motion with identity covariance. Using polar coordinates $(x,y)=(r \cos \theta,r\sin\theta)$ and setting, for $r\geq 0,\theta\in [0,2\pi[$,
 \begin{itemize}
 \item $e_{1} = A^{-1}(1,0) = (- \sin(\beta), \cos(\beta))$
 \item $e_{2} = A^{-1}(0,1) = (\cos(\beta), - \sin(\beta))$
 \end{itemize}
 then $$\phi(r,\theta):=p^{up}(r \ A^{-1}(\cos(\theta),\sin(\theta)) = p^{up}\left(\dfrac{r}{\cos^{2}(\beta) - \sin^{2}(\beta)}(\cos(\beta + \theta),\sin(\theta-\beta))\right) $$
 is a solution of the Dirichlet problem
$$ \dfrac{1}{r} \dfrac{\partial }{\partial r} ( r \dfrac{\partial \phi}{\partial r}) + \dfrac{1}{r^{2}} \dfrac{\partial^{2} \phi}{\partial \theta^{2}} = 0, \ \ on \ \ \lbrace (r,\theta), \ r > 0, \ \theta \in (0,\pi/2) \rbrace $$
in the  cone $C=\{ (r,\theta),\qquad r >  0, \theta\in ]-\beta,\frac{\pi}{2}-\beta[ \}$,
with the boundary conditions $ \phi(\mathbb{R} e_{1}) = 1$ and $\phi(\mathbb{R} e_{2}) = 0$.
A solution, which in this case does not depend on $r$, is given by
$$  \phi(r,\theta) = \dfrac{1}{\pi/2 + \arcsin \rho}(- \theta +\pi/2 + \arcsin(\rho)/2),  $$
where $\rho$ is the correlation coefficient between the bid and ask queues.
Using the results of \cite{yoshida99}, this problem has a unique  bounded solution so finally
\begin{equation*}
p^{up}(x,y) = \dfrac{1}{\pi/2 + \arcsin \rho}\left(\pi/2 + \arcsin(\rho)/2 - \arctan(\dfrac{\sin(\arctan(y/x) - \beta)}{\cos(\beta+ \arctan(y/x))})  \right).
\end{equation*}
\endproof
\begin{remark}
When $\sqrt{\lambda_{a}} v_{a} = \sqrt{\lambda_{b}} v_{b}$, the probability $p^{up}(x,y)$ only depends on the ration $y/x$ and on the correlation $\rho$
\begin{equation}
p_{\rm up}(x,y)=\frac{1}{2}- \dfrac{ \arctan(\sqrt{ \frac{1+\rho}{1-\rho}} \frac{y-x}{y+x} ) } {2\arctan(\sqrt{ \frac{1+\rho}{1-\rho} })}\label{pupW.eq},
\end{equation}
and when $\rho = 0$ (which is the case for some empirical examples, see  Section \ref{sec.Markovapproximation}),
\begin{equation*}
p^{up}(x,y) = \dfrac{2}{\pi} \arctan(\dfrac{y}{x}).
\end{equation*}
Figure \ref{graphpup} displays  the dependence of the uptick probability $p^{up}$ on the bid-ask imbalance variable $\theta=\arctan(y/x)$ for different values of $\rho$.
\end{remark}

\clearpage
\section{Appendix: Technical Proofs}
\subsection{A $J_1$-continuity property}
\begin{lemma}\label{lemma.continuity} Let $\tau:D([0,\infty),\mathbb{R}^2)\mapsto [0,\infty[
$ be the first exit time from the positive orthant. The map
\begin{eqnarray}
  G  : (D([0,\infty),\mathbb{R}^2),J_1) & \to & \mathbb{R}\\
  \omega & \rightarrow & 1_{\omega(\tau(\omega)).(0,1) \leq 0} . \nonumber
\end{eqnarray}
is continuous on the set $\{\omega\in
C([0,\infty), \mathbb{R}^{2} \backslash \{(0,0)\}), \qquad
\tau(\omega)<\infty\}.$
\end{lemma}
When $\tau(\omega)<\infty$, $G(\omega)=1$ indicates that $\omega$ first exits the orthant by crossing the $x$-axis.
To prove this property, first note  that
\begin{equation*}
C([0,\infty), \mathbb{R}^{2} \backslash \{(0,0)\}) = \bigcup_{n \geq 1} C_{0}([0,\infty), \mathbb{R}^{2} \backslash  B(0,1/n)).
\end{equation*}
Let $\omega_{0} \in C([0,\infty), \mathbb{R}^{2} \backslash \{(0,0)\})$. There exists $n \in \mathbb{N}$ such that $\omega_{0} \notin B(0,1/n)$. Let $\epsilon >0$ such that $\epsilon +  \eta_{\omega_{0}}(\epsilon) + \eta_{\omega_{0} \circ \lambda}(\epsilon) < 1/n$, where $\eta_{\omega}$ is the modulus of continuity of $\omega$.
Let  $\omega' \in D([0,\infty),\mathbb{R}^2)$ with $d_{J_{1}}(\omega_{0},\omega') \leq \epsilon$. There exists $\lambda: [0,T] \rightarrow [0,T]$ increasing such that:
\begin{equation*}
||\omega_{0} \circ \lambda - \omega ||_{\infty} \leq \epsilon \quad \text{and} \quad || \lambda - e ||_{\infty}.
\end{equation*}
Without loss of generality, one can also assume, by continuity of $\tau$ on $(D,J_{1})$, that
\begin{equation*}
| \tau(\omega_{0}) - \tau(\omega)| \leq \epsilon.
\end{equation*}
Now, we will show that $|\omega_{0}(\tau(\omega_{0})) - \omega'(\tau(\omega')) | \leq \epsilon +  \eta_{\omega_{0}}(\epsilon) + \eta_{\omega_{0} \circ \lambda}(\epsilon)$:
\begin{equation*}
|\omega_{0}(\tau(\omega_{0})) - \omega'(\tau(\omega')) | =| \omega_{0}(\tau(\omega_{0})) - \omega_{0} \circ \lambda(\tau(\omega')) +  \omega_{0} \circ \lambda(\tau(\omega')) -  \omega_{0} \circ \lambda(\tau(\omega_{0})) +  \omega_{0} \circ \lambda(\tau(\omega_{0})) - \omega'(\tau(\omega'))|,
\end{equation*}
therefore
\begin{align*}
|\omega_{0}(\tau(\omega_{0})) - \omega'(\tau(\omega')) | & \leq ||\omega_{0} \circ \lambda - \omega' ||_{\infty} + | \omega_{0} \circ \lambda(\tau(\omega')) -  \omega_{0} \circ \lambda(\tau(\omega_{0}))| + |\omega_{0} \circ \lambda(\tau(\omega_{0})) - \omega_{0}(\tau(\omega_{0}))| \\
 & \leq \epsilon + \eta_{\omega_{0}}(\epsilon) + \eta_{\omega_{0} \circ \lambda}(\epsilon).
\end{align*}
Since $\epsilon+ \eta_{\omega_{0}}(\epsilon) + \eta_{\omega_{0} \circ \lambda}(\epsilon) < 1/n $ and $\omega_{0} \notin B(0,1,n)$, $1_{\tau(\omega_{0}).(0,1) \leq 0}  = 1_{\tau(\omega').(0,1) \leq 0} $, which completes the proof of the continuity of the map $G$ on the space $C([0,\infty), \mathbb{R}^{2} \backslash \{(0,0)\})$.

\subsection{Continuity of  $\Psi$: proof of Theorem \ref{prop.psicontinuity}} \label{sec.psicontinuity}

To study the continuity of the map $\Psi$, we endow $D([0,\infty),\mathbb{R}^2)$ with Skorokhod's $J_1$  topology (see \cite{lindvall73,whitt80}). Let $\Lambda_T $ the set of continuous, increasing functions $\lambda: [0,T] \to [0,T]$  and $e$ the identical function on $[0,T]$. Recal that the following metric
\begin{equation*}
d_{J_{1}}(\omega_{1}, \omega_{2}) = \inf_{\lambda \in \Lambda} \left( ||\omega_{2} \circ \lambda - \omega_{1} ||_{\infty} + ||\lambda - e ||_{\infty}   \right).
\end{equation*}
defined for $\omega_{1}, \omega_{2} \in D([0,T],\mathbb{R}^2)$, induces the
 $J_{1}$ topology  on $D([0,T],\mathbb{R}^2) $, and
$\omega_{n}\to \omega $ in  $(D([0,\infty),\mathbb{R}^2),J_1)$ if for every continuity point $T$ of  $\omega,$ $\omega_{n}\to \omega $ in $(D([0,T],\mathbb{R}^2), J_1).$

The set $(\mathbb{R}_+^2)^\mathbb{^N}$  is endowed with the topology induced by 'cylindrical' semi-norms, defined as follows: for a sequence $(R^n)_{n\geq 1}$ in $(\mathbb{R}_+^2)^\mathbb{^N}$
$$ R^n \mathop{\to}^{n\to\infty} R\quad \iff \forall k\geq 1,\quad \sup\{|R^n_1-R_1|,...,|R^n_k-R_k|)\mathop{\to}^{n\to\infty} 0.$$
$D([0,\infty),\mathbb{R}^2)\times (\mathbb{R}_+^2)^\mathbb{^N} \times (\mathbb{R}_+^2)^\mathbb{^N}$ is then endowed with the corresponding product topology.
The goal of this section is to characterize the continuity set of the map
$$
  \Psi: D([0,\infty),\mathbb{R}^2)\times (\mathbb{R}_+^2)^\mathbb{^N} \times (\mathbb{R}_+^2)^\mathbb{^N}  \mapsto  D([0,\infty),\mathbb{R}_+^2)  $$
introduced in Definition \ref{def.regularization}.
Let us introduce
$C([0,\infty), \mathbb{R}^{2} \backslash \{(0,0)\})$ be the space of continuous planar paths avoiding the origin:
\begin{equation*}
C([0,\infty), \mathbb{R}^{2} \backslash \{(0,0)\}) = \bigcup_{n \geq 1} C_{0}([0,T], \mathbb{R}^{2} \backslash  B(0,1/n)).
\end{equation*}
\begin{lemma} \label{continuityPsi1}
Let $\omega\in C([0,\infty), \mathbb{R}^{2} \backslash \{(0,0)\})$.
Then the map
\begin{eqnarray}
  \Psi_{1}: D([0,\infty),\mathbb{R}^2)\times \mathbb{R}_+ \times \mathbb{R}_+& \to & D([0,\infty),\mathbb{R}_+^2)  \label{eq.Psi1}\\
  (\omega,R_{1},\tilde{R}_{1}) & \mapsto & \omega + 1_{[\tau(\omega),\infty)}\left( 1_{\sigma_{b}(\omega) = \tau(\omega)}(R_{1} - \omega_{\tau(\omega)}) + 1_{\sigma_{a}(\omega) = \tau(\omega)}(\tilde{R}_{1} - \omega_{\tau(\omega)} ) \right) \nonumber,
\end{eqnarray}
where
\begin{equation*}
\sigma_{b}(\omega) = \inf \lbrace t \geq 0, \  \omega_{t}.(0,1) \leq 0  \rbrace, \quad \sigma_{a}(\omega) = \inf \lbrace t \geq 0, \  \omega_{t}.(1,0) \leq 0  \rbrace \quad \text{and} \quad \tau(\omega) = \sigma_{b}(\omega) \wedge \sigma_{a}(\omega).
\end{equation*}
is continuous   at $\omega$
with respect to the following distance on
$(D([0,\infty],\mathbb{R}^{2}) \times \mathbb{R}_+ \times \mathbb{R}_+)$:
\begin{equation*}
d((\omega,R_{1},\tilde{R}_{1}),(\omega',R'_{1},\tilde{R}'_{1})) = d_{J_{1}}(\omega,\omega') + |R_{1} -R_{1}'| + |\tilde{R}_{1} - \tilde{R}_{1}| \quad \text{and}:
\end{equation*}
\end{lemma}
\proof{Proof.}
Let $(\omega_{0},R_{1},\tilde{R}_{1}) \in C([0,\infty), \mathbb{R}^{2} \backslash \{(0,0)\})\times \mathbb{R}_{+}^2$, $(\omega',R'_{1},\tilde{R}'_{1}) \in D(0,\mathbb{R}^{2}) \times \mathbb{R}_{+}^2$.
 Since $\omega_{0} \in C([0,\infty), \mathbb{R}^{2} \backslash \{(0,0)\})$, there exists $n >0$ such that $\omega_{0} \notin B(0,1/n)$. Let $0 < \epsilon < 1/n $ such that
\begin{equation*}
d((\omega_{0},R_{1},\tilde{R}_{1}),(\omega',R'_{1},\tilde{R}'_{1})) < \epsilon.
\end{equation*}
Since $d_{J_{1}}(\omega_{0},\omega') < \epsilon$, there exists $\lambda : [0,T] \rightarrow [0,T]$, non-decreasing such that:
\begin{equation*}
|| \lambda - e ||_{\infty} < \epsilon, \quad \text{and} \quad || \omega_{0} \circ \lambda - \omega ||_{\infty} < \epsilon.
\end{equation*}

By continuity of $\tau$ for the $J_1$ topology \cite{Whitt}[Theorem 13.6.4], one can also assume, without loss of generality, that
\begin{equation*}
|\tau(\omega_{0} \circ \lambda) - \tau(\omega') | \leq \epsilon.
\end{equation*}
Moreover, since the graph of $\omega_{0}$  does not intersect with $\overline{B(0,1/n)}$ and $\epsilon < 1/n$, $1_{\tau(\omega_{0}) = \sigma_{a}(\omega_{0})} = 1_{\tau(\omega') = \sigma_{a}(\omega')}$.
Now define $\lambda^{\epsilon}$ by
\begin{eqnarray}
 \lambda^{\epsilon}: [0,T]& \to & [0,T]  \\
 t & \mapsto & \dfrac{\tau(\omega')}{\tau(\omega_{0} \circ \lambda)} \lambda_{t} \nonumber.
\end{eqnarray}
Then
\begin{eqnarray*}
||\lambda^{\epsilon} -e||_{\infty} =  ||\frac{\tau(\omega)}{\tau(\omega_{0} \circ \lambda)} \lambda -e||_{\infty}  \leq ||\frac{\tau(\omega)}{\tau(\omega_{0} \circ \lambda)} \lambda - \frac{\tau(\omega)}{\tau(\omega_{0} \circ \lambda)}e ||_{\infty}+ || \frac{\tau(\omega)}{\tau(\omega_{0} \circ \lambda)} e  -e||_{\infty} \\
\leq \epsilon \frac{\tau(\omega)}{\tau(\omega_{0} \circ \lambda)}+ \frac{\epsilon}{\tau(\omega_{0} \circ \lambda)}.
\end{eqnarray*}
On the other hand
\begin{equation*}
||\omega_{0} \circ \lambda^{\epsilon}  - \omega ||_{\infty} = ||\omega_{0} \circ \lambda^{\epsilon} - \omega_{0} \circ \lambda + \omega_{0} \circ \lambda  - \omega ||_{\infty} \leq ||\omega_{0} \circ \lambda^{\epsilon} - \omega_{0} \circ \lambda ||_{\infty}  + \epsilon \leq \eta_{\omega_{0} \circ \lambda} (\epsilon) + \epsilon,
\end{equation*}
where $\eta_{\omega_{0} \circ \lambda}$ is the modulus of continuity modulus of  $\omega_{0} \circ \lambda$.
Therefore, since$1_{\tau(\omega_{0} \circ \lambda^{\epsilon})} = 1_{\tau(\omega')}$  by definition of $\lambda^{\epsilon}$ and
\begin{align*}
\Psi_{1}(\omega_{0},R_{1},\tilde{R}_{1})\circ \lambda^{\epsilon} - \Psi_{1}(\omega',R'_{1},\tilde{R}'_{1})) & = \omega_{0} \circ \lambda^{\epsilon} - \omega' \\
 &+ 1_{\tau(\omega_{0} \circ \lambda^{\epsilon})}\left( 1_{\tau(\omega') = \sigma_{a}} (R'_{1} - R_{1}) + 1_{\tau(\omega') = \sigma_{b}} (\tilde{R}'_{1} - \tilde{R}_{1}) \right).
\end{align*}
Thus $\lambda^{\epsilon}$ satisfies $|| \lambda^{\epsilon}  - e|| \leq \epsilon(\frac{\tau(\omega') +1}{\tau(\omega_{0} \circ \lambda )})$ and
\begin{equation*}
|| \Psi_{1}(\omega_{0} ,R_{1},\tilde{R}_{1})\circ \lambda^{\epsilon} - \Psi_{1}(\omega',R'_{1},\tilde{R}'_{1})) ||_{\infty} \leq \eta_{\omega_{0} \circ \lambda} (\epsilon) + \epsilon + 2 \epsilon
\end{equation*}
which proves that $(\omega_{0} ,R_{1},\tilde{R}_{1})$ is a continuity point for $\Psi_{1}$.
\endproof
For $k\geq 2$, define recursively the maps
\begin{eqnarray}
  \Psi_{k}: D([0,\infty),\mathbb{R}^2)\times \mathbb{R}_+^\mathbb{N} \times \mathbb{R}_+^\mathbb{N} & \to & D([0,\infty),\mathbb{R}_+^2)  \label{eq.Psik}\\
  (\omega,(R_{i},\tilde{R}_{i})_{i \geq 1}) & \mapsto & \Psi_{1}(\Psi_{k-1}(\omega,(R_i,\tilde{R}_{i})_{i =1..k-1}),R_{k},\tilde{R}_{k}). \nonumber
\end{eqnarray}
To simplify notation we will denote the argument of $\Psi_{k}$ as $ (\omega,R,\tilde{R})(=(\omega,(R_{i},\tilde{R}_{i})_{i \geq 1})$ although it is easily observed from  \eqref{eq.Psik} that $\Psi_k$ only depends on the first $k$ elements  $(R_i,\tilde{R}_{i})_{i =1..k})$ of  $R,\tilde{R}$.
\begin{lemma}\label{continuitypsik}
If $(\omega, R, \tilde{R})\in C([0,\infty), \mathbb{R}^{2} \backslash \{(0,0)\})\times \mathbb{R}_+^\mathbb{N} \times \mathbb{R}_+^\mathbb{N}$ such that
\begin{equation}
(0,0)\notin \Psi_k(\omega,R,\tilde{R})( [0,\infty)\ )
\end{equation}
then $\Psi_k$ is continuous at $(\omega, R, \tilde{R})$.
\end{lemma}
\proof{Proof.}
Let $(R_{i},\tilde{R}_{i})_{i \geq 1}$, $(R'_{i},\tilde{R}'_{i})_{i \geq 1}$, two sequences of random variables on $\mathbb{R}_+^{2}$ and define
 \begin{equation*}\Omega_{k}(R,\tilde{R}) =  \cap_{j=0}^k \Psi_{j}(C([0,\infty), \mathbb{R}^{2} \backslash \{(0,0)\}),R,\tilde{R})
\end{equation*}
where we have set $\Psi_0=Id$. Consider $\omega_{0} \in \Omega_{k}(R,\tilde{R})$, and $\omega \in D([0,T], \mathbb{R}_+^{2})$, such that:
\begin{equation*}
d_{J_1}(\omega_{0},\omega)+ \sup_{i=1..k} |R_{i}-R'_{i}|+ \sup_{i=1..k} |\tilde{R}_{i}-\tilde{R}'_{i}|\leq \epsilon.
\end{equation*}
An application of  the triangle inequality yields
\begin{align*}
 d_{J_{1}}(\Psi_{k}(\omega_{0},(R_{i},\tilde{R}_{i})),\Psi_{k}(\omega',(R'_{i},\tilde{R}'_{i}))) \\
  \leq
     d_{J_{1}}(\Psi_{k}(\omega_{0},(R_{i},\tilde{R}_{i})),\Psi_{k}(\omega',(R_{i},\tilde{R}_{i})))
  +  d_{J_{1}}(\Psi_{k}(\omega',(R_{i},\tilde{R}_{i})),\Psi_{k}(\omega',(R'_{i},\tilde{R}'_{i})))
\end{align*}
where the last term converges to zero when $\epsilon$ goes to zero by continuity of $\Psi_{1}$.
\endproof
We can now prove Theorem \ref{prop.psicontinuity}.
\proof{Proof of Theorem \ref{prop.psicontinuity}.}
Since $\omega$ is continuous, the jumps of $\Psi(\omega,R,\tilde{R})$ correspond to the first exit times from the orthant of the paths $\Psi_k(\omega,R,\tilde{R})$.
Therefore, if $(R_n)_{n\geq 1},(\tilde{R}_n)_{n\geq 1}$ have no accumulation points on the axes, the paths $\Psi(\omega,R,\tilde{R})$ only has  a finite number of discontinuities on $[0,T]$ for any $T>0$. So, for any  $T>0$, there exists $k(T)$ such that $\Psi = \Psi_{k(T)}$. Then thanks to Lemma \ref{continuitypsik}, $\Psi$ is continuous on the set of continuous trajectories whose image has  a finite number of discontinuities and does not contain the origin.
\endproof
\subsection{Functional central limit theorem for the net order flow}

\begin{proposition} \label{Diffusion-thm}
Let
$(T_{i}^{a,n},T_{i}^{b,n})_{i \geq 1}$ and $(V_{i}^{a,n},V_{i}^{b,n})_{i \geq 1}$ be stationary arrays of random variables  which satisfy Assumptions \ref{assumption.G} and \ref{assumption.H}.
Let
$(N_{t}^{a,n}, t \geq 0)$ and $(N_{t}^{b,n}, t \geq 0)$ be the counting processes defined in \eqref{eq.NaNb}.
Then
\begin{equation} \label{diffusive}
\left( \sum_{i=1}^{ N^{a,n}_{nt}}\dfrac{ V_{i}^{a,n}}{\sqrt{n}},
\sum_{i=1}^{ N^{b,n}_{nt}} \dfrac{V_{i}^{b,n}}{\sqrt{n}} \right)_{t
\geq 0} \mathop{\Rightarrow}_{n \rightarrow \infty}^{J_1} \left(
\Sigma B_{t} + t(\lambda^{a} \overline{V^{a}},\lambda^{b}\overline{V^{b}}) \right)_{t
\geq 0}
\end{equation}
where $B$ is a standard planar Brownian motion and
\begin{equation}
 \Sigma {}^t \! \Sigma=   \begin{pmatrix}
\lambda^{a} v_{a}^{2} & \rho  \sqrt{\lambda^{a} \lambda^{b}} v_{a} v_{b} \\
\rho  \sqrt{\lambda^{a} \lambda^{b}} v_{a} v_{b} & \lambda^{b}
v_{b}^{2}
\end{pmatrix} ,
\end{equation}
\end{proposition}
\textbf{Proof:}
First we will prove that the sequence of processes
\begin{equation*} \label{convergenceCW}
\left( \sum_{i=1}^{ [\lambda^{a}t]}\dfrac{ V_{i}^{a,n}}{\sqrt{n}},
\sum_{i=1}^{ [\lambda^{b}t]} \dfrac{V_{i}^{b,n}}{\sqrt{n}}
\right)_{t \geq 0} \mathop{\Rightarrow}_{n \rightarrow \infty}^{J_1}
\left( \Sigma B_{t} + t(\lambda^{a} \overline{V^{a}},\lambda^{b}\overline{V^{b}})
\right)_{t \geq 0}
\end{equation*}
weakly converges in the $J_1$ topology.  Using  the
Cramer-Wold device, it is sufficient to prove that for $(\alpha,\beta) \in \mathbb{R}^{2}$,
\begin{equation*}
\left( \alpha \sum_{i=1}^{ [\lambda^{a}t]}\dfrac{
V_{i}^{a,n}}{\sqrt{n}} + \beta \sum_{i=1}^{ [\lambda^{b}t]}
\dfrac{V_{i}^{b,n}}{\sqrt{n}} \right)_{t \geq 0} \mathop{\Rightarrow}_{n \rightarrow \infty}
(\alpha \lambda^{a} \overline{V^{a}}+ \beta\lambda^{b}\overline{V^{b}}))t +
\sqrt{(\alpha^{2} \lambda^{a} v_{a}^{2} + \beta^{2} \lambda^{b}
v_{b}^{2} + 2 \rho \alpha \beta v_{a} v_{b} \sqrt{ \lambda^{a}
\lambda^{b}} )}B_{t}
\end{equation*}
If $\lambda^{a} \in \mathbb{Q}$ and $\lambda^{b} \in \mathbb{Q}$, it
is possible to find $\lambda$ such that $\lambda^{a}/\lambda \in
\mathbb{N}$ and $\lambda^{b} / \lambda \in \mathbb{N}$. Let for all
$(i,n) \in \mathbb{N}^{2}$,
\begin{equation*}
W_{i}^{n} = \alpha \left( V_{(\lambda^{a}/\lambda)(i-1) + 1}^{a,n} +
V_{2}^{a,n} + ... + V^{a,n}_{\lambda^{a}i/\lambda} \right) + \beta
\left( V_{(\lambda^{b}/\lambda)(i-1) + 1}^{b,n} + V_{2}^{b,n} + ...
+ V^{b,n}_{\lambda^{b}i/\lambda} \right),
\end{equation*}
then for all $t > 0$,
\begin{equation*}
\alpha \sum_{i=1}^{ [\lambda^{a}t]}\dfrac{ V_{i}^{a,n}}{\sqrt{n}} +
\beta \sum_{i=1}^{ [\lambda^{b}t]} \dfrac{V_{i}^{b,n}}{\sqrt{n}} =
\sum_{i=1}^{[\lambda t]} \dfrac{W_{i}^{n}}{\sqrt{n}}.
\end{equation*}
For all $n > 0$, $(W_{i}^{n}, i \geq 1)$ is a sequence of stationary
random variables. Therefore, thanks to theorem \cite[Chap.VIII, Thm
2.29, p.426]{jacodshiryaev}, and the fact that
\begin{equation} \label{conv}
{\rm var}(W^{n}_{1}) + 2 \sum_{i =2}^{\infty} cov(W^{n}_{1},W^{n}_{i})
\mathop{\rightarrow}^{n \rightarrow \infty} \sigma^2,
\end{equation}
the sequence of processes $\left( \sum_{i=1}^{[\lambda n t]}
\dfrac{W_{i}^{n}}{\sqrt{n}}, \ t \geq 0 \right)_{n \geq 1}$
converges weakly to a Brownian motion with volatility $\sqrt{\lambda}
\sigma$.
If $(\lambda^{a},\lambda^{b})\notin \mathbb{Q}^2$, there exists
$(\lambda_{n}^{a},\lambda_{n}^{b})_{ n \geq 1}$ such that
\begin{equation*}
\lambda_{n}^{a}, \lambda_{n}^{b}  \in \mathbb{Q} \quad  and \ \
|\lambda_{n}^{a} - \lambda^{a}| \leq \frac{1}{n}, \quad |\lambda_{n}^{b} - \lambda^{b}| \leq \frac{1}{n}.
\end{equation*}
As above, one can define an integer $\lambda_{n}$ such
that $\frac{\lambda^{a}_{n}}{\lambda_{n}} \in \mathbb{Q}$ and
$\frac{\lambda^{b}_{n}}{\lambda_{n}} \in \mathbb{Q}$. Let for all
$(i,n) \in \mathbb{N}^{2}$,
\begin{equation*}
W_{i}^{n} = \alpha \left( V_{(\lambda_{n}^{a}/\lambda_{n})(i-1) +
1}^{a,n} + V_{2}^{a,n} + ... +
V^{a,n}_{\lambda_{n}^{a}i/\lambda_{n}} \right) + \beta \left(
V_{(\lambda_{n}^{b}/\lambda_{n})(i-1) + 1}^{b,n} + V_{2}^{b,n} + ...
+ V^{b,n}_{\lambda_{n}^{b}i/\lambda_{n}} \right),
\end{equation*}
One has for all $t > 0$,
\begin{equation*}
\alpha \sum_{i=1}^{ [\lambda^{a}t]}\dfrac{ V_{i}^{a,n}}{\sqrt{n}} +
\beta \sum_{i=1}^{ [\lambda^{b}t]} \dfrac{V_{i}^{b,n}}{\sqrt{n}} =
\sum_{i=1}^{[\lambda_{n} t]} \dfrac{W_{i}^{n}}{\sqrt{n}} + \alpha
\sum_{i=1}^{ [\lambda^{a}t- \lambda_{n}^{a}t]}\dfrac{
V_{i}^{a,n}}{\sqrt{n}} + \beta \sum_{i=1}^{ [\lambda^{b}t-
\lambda_{n}^{b}t]} \dfrac{V_{i}^{b,n}}{\sqrt{n}} .
\end{equation*}
Moreover
\begin{equation*}
\left( \alpha \sum_{i=1}^{ [\lambda^{a}t- \lambda_{n}^{a}t]}\dfrac{
V_{i}^{a,n}}{\sqrt{n}} + \beta \sum_{i=1}^{ [\lambda^{b}t-
\lambda_{n}^{b}t]}   \dfrac{V_{i}^{b,n}}{\sqrt{n}} \right)_{t \geq
0} \Rightarrow^{J_1}  0,
\end{equation*}
therefore the convergence above holds even if $\lambda_{a}$ or
$\lambda_{b}$ are not rationals.
On one hand,
\begin{align*}
{\rm var}(W_{i}^{n})  & = & {\rm var} \left(\alpha ( V_{(\lambda_{n}^{a}/\lambda_{n})(i-1) +
1}^{a,n} + ... +
V^{a,n}_{\lambda_{n}^{a}i/\lambda_{n}} ) + \beta (
V_{(\lambda_{n}^{b}/\lambda_{n})(i-1) + 1}^{b,n} + V_{2}^{b,n} + ...
+ V^{b,n}_{\lambda_{n}^{b}i/\lambda_{n}} ) \right) \\
& = & \alpha^{2} {\rm var} \left( V_{(\lambda_{n}^{a}/\lambda_{n})(i-1) +
1}^{a,n} ... + V^{a,n}_{\lambda_{n}^{a}i/\lambda_{n}} ) \right) +  \beta^{2} {\rm var} \left( V_{(\lambda_{n}^{b}/\lambda_{n})(i-1) + 1}^{b,n} ... + V^{b,n}_{\lambda_{n}^{b}i/\lambda_{n}} ) \right) \\
 & + & 2 \alpha \beta{\rm cov}\left( V_{(\lambda_{n}^{a}/\lambda_{n})(i-1) +
1}^{a,n} ... +
V^{a,n}_{\lambda_{n}^{a}i/\lambda_{n}}, V_{(\lambda_{n}^{b}/\lambda_{n})(i-1) + 1}^{b,n} ... + V^{b,n}_{\lambda_{n}^{b}i/\lambda_{n}} ) \right). \\
\end{align*}

On the other hand, for all $i \geq 2$,

\begin{align*}
{\rm cov}(W_{1}^{n}, W_{i}^{n}) & = & \alpha^{2}{\rm cov}\left(V_{1}^{a,n} + ... + V_{(\lambda_{n}^{a}/\lambda_{n})}^{a,n}  , V_{(\lambda_{n}^{a}/\lambda_{n})(i-1) +
1}^{a,n} + ... +
V^{a,n}_{\lambda_{n}^{a}i/\lambda_{n}}\right)  \\
 & + &  \beta^{2}{\rm cov}\left(V_{1}^{b,n} + ... + V_{(\lambda_{n}^{b}/\lambda_{n})}^{b,n}  , V_{(\lambda_{n}^{b}/\lambda_{n})(i-1) +
1}^{b,n} + ... +
V^{b,n}_{\lambda_{n}^{b}i/\lambda_{n}}\right) \\
& + & \alpha \beta{\rm cov}\left( V_{1}^{a,n} + ... + V_{(\lambda_{n}^{a}/\lambda_{n})}^{a,n}, V_{(\lambda_{n}^{b}/\lambda_{n})(i-1) +
1}^{b,n} + ... +
V^{b,n}_{\lambda_{n}^{b}i/\lambda_{n}} \right) \\
& + & \alpha \beta{\rm cov}\left(V_{1}^{b,n} + ... + V_{(\lambda_{n}^{b}/\lambda_{n})}^{b,n} , V_{(\lambda_{n}^{a}/\lambda_{n})(i-1) +
1}^{a,n} + ... +
V^{a,n}_{\lambda_{n}^{a}i/\lambda_{n}} \right).
\end{align*}

Therefore
\begin{align*}
{\rm var} (W_{1}^{n}) + 2 \sum_{i=2}^{\infty}
{\rm cov} (W_{1}^{n},W_{i}^{n})
& = & {\rm var} ( V_{1}^{a,n}) \frac{\lambda_{n}^{a}}{\lambda_{n}} + 2 \sum_{i = 2}^{\infty}{\rm cov}(V_{1}^{a,n}, V_{i}^{a,n}) \frac{\lambda_{n}^{a}}{\lambda_{n}} \\
& + &  {\rm var} ( V_{1}^{b,n}) \frac{\lambda_{n}^{b}}{\lambda_{n}} + 2 \sum_{i = 2}^{\infty}{\rm cov}(V_{1}^{b,n}, V_{i}^{b,n}) \frac{\lambda_{n}^{b}}{\lambda_{n}} \\
& + & 2 \alpha \beta{\rm cov}\left( V_{1}^{a,n} ... +
V^{a,n}_{\lambda_{n}^{a}/\lambda_{n}}, V_{1}^{b,n} ... + V^{b,n}_{\lambda_{n}^{b}/\lambda_{n}}  \right) \\
&+ & 2 \alpha \beta \sum_{i= 2}^{\infty}{\rm cov}\left( V_{1}^{a,n} ... +
V^{a,n}_{\lambda_{n}^{a}/\lambda_{n}},  V_{(\lambda_{n}^{b}/\lambda_{n})(i-1) +
1}^{b,n} + ... +
V^{b,n}_{\lambda_{n}^{b}i/\lambda_{n}} \right) \\
&+& 2 \alpha \beta \sum_{i= 2}^{\infty}{\rm cov}\left(V_{1}^{b,n} ... + V^{b,n}_{\lambda_{n}^{b}/\lambda_{n}}, V_{(\lambda_{n}^{a}/\lambda_{n})(i-1) +
1}^{a,n} + ... +
V^{a,n}_{\lambda_{n}^{a}i/\lambda_{n}}   \right)
\end{align*}
 A simple calculation shows that
\begin{align*}
 2 \alpha \beta {\rm cov}\left( V_{1}^{a,n} ... +
V^{a,n}_{\lambda_{n}^{a}/\lambda_{n}}, V_{1}^{b,n} ... + V^{b,n}_{\lambda_{n}^{b}/\lambda_{n}}  \right) \\
&+ & 2 \alpha \beta \sum_{i= 2}^{\infty}{\rm cov}\left( V_{1}^{a,n} ... +
V^{a,n}_{\lambda_{n}^{a}/\lambda_{n}},  V_{(\lambda_{n}^{b}/\lambda_{n})(i-1) +
1}^{b,n} + ... +
V^{b,n}_{\lambda_{n}^{b}i/\lambda_{n}} \right) \\
&+& 2 \alpha \beta \sum_{i= 2}^{\infty}{\rm cov}\left(V_{1}^{b,n} ... + V^{b,n}_{\lambda_{n}^{b}/\lambda_{n}}, V_{(\lambda_{n}^{a}/\lambda_{n})(i-1) +
1}^{a,n} + ... +
V^{a,n}_{\lambda_{n}^{a}i/\lambda_{n}}   \right) \\
 =  2 \alpha \beta {\rm max}( \frac{\lambda_{n}^{a}}{\lambda_{n}},\frac{\lambda_{n}^{b}}{\lambda_{n}}){\rm cov}(V_{1}^{a,n},V_{1}^{b,n})
& +& 2 \alpha \beta \sum_{i=2}^{\infty}
\frac{\lambda_{n}^{a}}{\lambda_{n}}{\rm cov}(V_{1}^{a,n},V_{i}^{b,n}) + \frac{\lambda_{n}^{b}}{\lambda_{n}}
{\rm cov}(V_{1}^{b,n},V_{i}^{a,n}).
\end{align*}

Therefore
\begin{equation*}
\lim_{n \mapsto \infty}  {\rm var}(W_{1}^{n}) + 2 \sum_{i=2}^{\infty}
{\rm cov}(W_{1}^{n},W_{i}^{n}) = \alpha \dfrac{\lambda^{a}}{\lambda}
v_{a}^{2} + \beta \dfrac{\lambda^{b}}{\lambda} v_{b}^{2} + 2 \rho
\sqrt{\alpha \beta } \dfrac{\sqrt{\lambda^{a} \lambda^{b}}}{\lambda}
v_{a} v_{b},
\end{equation*}
where $\rho$ is given in \eqref{eq.rho} and
\begin{equation*}
\lim_{n \mapsto \infty}  \mathbb{E}[W_{i}^{n}] = \alpha
\dfrac{\lambda^{a}}{\lambda} \overline{V^a} + \beta
\dfrac{\lambda^{b}}{\lambda} \overline{V^b},
\end{equation*}
which completes the proof of the convergence in \eqref{conv}. The
law of large numbers for renewal processes implies that the
following sequence of processes converges to zero in the $J_1$ topology \cite{iglehart71}:
\begin{equation*}
(N_{nt}^{a,n})_{t \geq 0}   \mathop{\Rightarrow}^{n \rightarrow \infty}
([\lambda^{a} t])_{t \geq 0}, \ \ and \ \  (N_{nt}^{b,n})_{t \geq 0}
\mathop{\Rightarrow}^{n \rightarrow \infty}  ([\lambda^{b} t])_{t \geq 0},
\end{equation*}
\begin{equation*}
\left( \sum_{i=[\lambda^{a}t]}^{N_{nt}^{a,n}}
\dfrac{V_{i}^{a,n}}{\sqrt{i}},
\sum_{i=[\lambda^{b}t]}^{N_{nt}^{b,n}} \dfrac{V_{i}^{b,n}}{\sqrt{i}}
\right)_{t \geq 0} \Rightarrow 0\qquad{\rm in\ the\ }J_1\ {\rm topology}.
\end{equation*}

\subsection{Identification of the heavy traffic limit $Q$}\label{appendix.generator}
\begin{lemma}\label{lemma.generator}
The process $Q$ is a Markov process with values in $\mathbb{R}_{+}^{2}-\{(0,0)\}$ and infinitesimal generator given by \eqref{Generator}- \eqref{eq.boundary}
and domain
\begin{eqnarray*}{\rm dom}(\mathcal{G})= \{ h \in {C}^{2}(]0,\infty[\times]0,\infty[,\mathbb{R})\cap {C}^{0}(\mathbb{R}_{+}^{2},\mathbb{R}),&\qquad \forall x>0,\qquad \forall y>0,\\
h(x,0) =  \int_{\mathbb{R}_{+}^2} h(g((x,0),(u,v))) F(du,dv)=0,&\quad
h(0,y) = \int_{\mathbb{R}_{+}^2} h(g((0,y),(u,v))) \tilde{F}(du,dv)=0 \}
\end{eqnarray*}
\end{lemma}
\proof{Proof.}
We use the explicit construction of $Q$ from the planar Brownian motion $X$ using the maps $\Psi_k$ defined in the proof of Theorem \ref{HeavyTraffic.thm}.
First, let us show that the process $Q$ is a Markov process. Let $0 \leq t_{1} < ... < t_{n} \leq t$, and $k$ such that $\tau_{k} \leq t_{n} \leq \tau_{k+1}$. Let $A$ a measurable set. The following equations:
\begin{align*}
\mathbb{P}[Q_{t} \in  A  | Q_{t_{1}}, ..., Q_{t_{n}} ] & = \mathbb{P}[Q_{t} \in  A  | Q_{t_{1}}, ..., Q_{t_{n}}, \lbrace t_{n} < t < \tau_{k+1} \rbrace] \mathbb{P}[t_{n} < t < \tau_{k+1} | Q_{t_{1}}, ..., Q_{t_{n}} ] \\
 & +  \mathbb{P}[Q_{t} \in  A  | Q_{t_{1}}, ..., Q_{t_{n}}, \lbrace t \geq \tau_{k+1} \rbrace] \mathbb{P}[t \geq \tau_{k+1} | Q_{t_{1}}, ..., Q_{t_{n}} ] \\
 & = \mathbb{P}[Q_{t} \in  A  | Q_{t_{n}}, \lbrace t_{n} < t < \tau_{k+1} \rbrace] \mathbb{P}[t_{n} < t < \tau_{k+1} | Q_{t_{n}} ] \\
 & +  \mathbb{P}[Q_{t} \in  A  | Q_{t_{n}}, \lbrace t \geq \tau_{k+1} \rbrace]  \mathbb{P}[t \geq \tau_{k+1} | Q_{t_{n}} ] \\
 & = \mathbb{P}[Q_{t} \in  A  | Q_{t_{n}} ]
\end{align*}
prove that the process $Q$ is a Markov process. Let us now compute its infinitesimal generator $\mathcal{G}$.
The domain  ${\rm dom}(\mathcal{G})$ of  $\mathcal{G}$ consists of all functions $h \in \mathcal{C}^{2}(\mathbb{R}_{+}^{2})$ verifying for all $(x,y) \in \mathbb{R}_{+}^{2}$
\begin{equation*}
\lim_{t \rightarrow 0} \dfrac{\mathbb{E}[h(Q_{t})-h(Q_{0}) | Q_{0} = (x,y)]}{t} < \infty.
\end{equation*}
For $x >0$, and $y > 0$, a  classical computation shows that if $h \in \mathcal{C}^{2}(\mathbb{R}_{+}^{2})$,
\begin{equation*}
\mathbb{E}[h(Q_{t}) | Q_{0} = (x,y)] = h(x,y) + t \left(  \lambda_{a} \overline{V^a} \dfrac{\partial h}{\partial x} +  \lambda_{b} \overline{V^b} \dfrac{\partial h}{\partial y} + \dfrac{ \lambda_{a} v_{a}^{2}}{2} \dfrac{\partial^{2} h}{\partial x^{2}} + \dfrac{ \lambda_{b} v_{b}^{2}}{2} \dfrac{\partial^{2} h}{\partial y^{2}} + 2 \rho \sqrt{\lambda_{a} \lambda_{b}} v_{a} v_{b} \dfrac{\partial^{2} h}{\partial x \partial y}  \right) + o(t),
\end{equation*}
which leads to equation \eqref{Generator}. On the other hand, for all $(x,y) \in \mathbb{R}_{+}^{2}$,
\begin{align*}
\mathbb{E}[h(Q_{t}) | Q_{0} = (x,0)] & = \int_{\mathbb{R}_+^2} \mathbb{E}[h(Q_{t}) | Q_{0+} = g((x,0),(u,v))] F(du,dv) \\
 &=  \int_{\mathbb{R}_+^2} \left( \mathbb{E}[h(Q_{t}) | Q_{0+} = g((x,0),(u,v))] - h(g((x,0),(u,v))) + h(g((x,0),(u,v)))  \right) F(du,dv) \\
 & = \int_{\mathbb{R}_+^2} \left( t \mathcal{G}h(g((x,0),(u,v)))  + h(g((x,0),(u,v)))  \right) F(du,dv) + o(t).
\end{align*}

\begin{align*}
\mathbb{E}[h(Q_{t}) | Q_{0} = (0,y)] & = \int_{\mathbb{R}_+^2} \mathbb{E}[h(Q_{t}) | Q_{0+} = g((0,y),(u,v))] \tilde{F}(du,dv) \\
 &=  \int_{\mathbb{R}_+^2} \left( \mathbb{E}[h(Q_{t}) | Q_{0+} = g((0,y),(u,v))] - h(g((0,y),(u,v))) + h(g((0,y),(u,v)))  \right) \tilde{F}(du,dv) \\
 & = \int_{\mathbb{R}_+^2} \left( t \mathcal{G}h(g((0,y),(u,v)))  + h(g((0,y),(u,v)))  \right) \tilde{F}(du,dv) + o(t).
\end{align*}
As $t\to 0$, these equations lead to\begin{align*}
\dfrac{\mathbb{E}[h(Q_{t}) | Q_{0} = (x,0)] - h(x,0)}{t} & = \int_{\mathbb{R}_+^2} \mathcal{G}h(g((x,0),(u,v))) F(du,dv)  \\
& + \frac{1}{t} \int_{\mathbb{R}_+^2} h(g((x,0),(u,v))) - h(x,0) F(du,dv) + o(1).
\end{align*}
\begin{align*}
\dfrac{\mathbb{E}[h(Q_{t}) | Q_{0} = (0,y)] - h(0,y)}{t} & = \int_{\mathbb{R}_+^2} \mathcal{G}h(g((0,y),(u,v))) \tilde{F}(du,dv) \\
 & + \frac{1}{t} \int_{\mathbb{R}_+^2} \left(h(g((0,y),(u,v))) - g(0,y) \right) \tilde{F}(du,dv) + o(1).
\end{align*}
Therefore, the domain ${\rm dom}(\mathcal{G})$ of $\mathcal{G}$ consists of all function $h \in \mathcal{C}^{2}(\mathbb{R}_{+}^{*} \times \mathbb{R}_{+}^{*})$ such that
\begin{equation*}
h(x,0) =  \int_{\mathbb{R}_{+}^2} h(g((x,0),(u,v))) F(du,dv)=0,\qquad
h(0,y) = \int_{\mathbb{R}_{+}^2} h(g((0,y),(u,v))) \tilde{F}(du,dv)=0,
\end{equation*}
When these 'boundary conditions' are verified, the above limits exist and we have
$$ \mathcal{G}h(x,0) =\int_{\mathbb{R}_+^2} \mathcal{G} h(g((x,0),(u,v))) F(du,dv) ,    $$
$$ \mathcal{G}h(0,y) = \int_{\mathbb{R}_+^2} \mathcal{G} h(g((0,y),(u,v))) \tilde{F}(du,dv),   $$
which concludes the proof.
\endproof

\end{document}